\documentclass[11pt]{article}

\usepackage{amssymb,latexsym,amsmath}

\usepackage{fullpage}
\usepackage{nicefrac}
\usepackage[pdftex]{graphicx}
\usepackage{colortbl}
\usepackage{mathtools}
\usepackage{wasysym}
\usepackage{caption,subcaption,wrapfig}
\usepackage[colorlinks=true,linktocpage=true,linkcolor=blue,citecolor=blue]{hyperref}

\usepackage{tikz}
\usetikzlibrary{decorations.markings,decorations.pathmorphing}

\definecolor{gray}{rgb}{.9,.9,.9}

\newcommand{\dd}{\mathrm{d}}

\newcommand{\eqq}[1]{(\ref{#1})}
\newcommand{\cpthree}{\mathbb{CP}^3}
\newcommand{\Fig}[1]{Fig.~\ref{#1}}
\newcommand{\Sec}[1]{Sec.~\ref{#1}}

\newcommand{\be}{\begin{equation}}
\newcommand{\ee}{\end{equation}}
\newcommand{\sac}{\, , \qquad}

\newcommand{\Bconf}{\mathbb{B}_8^{\rm{conf}}}
\newcommand{\Bop}{\mathbb{B}_8^{\textrm{\tiny OP}}}
\newcommand{\Bplus}{\mathbb{B}_8^+}
\newcommand{\Bminus}{\mathbb{B}_8^-}
\newcommand{\Binf}{\mathbb{B}_8^\infty}

\definecolor{gray}{rgb}{.9,.9,.9}

\makeatletter \@addtoreset{equation}{section}
\makeatother

\begin{document}
\begin{titlepage}
	\thispagestyle{empty}
	\begin{flushright}
		\hfill{ICCUB-17-005}
	\end{flushright}
	
	\vspace{55pt}  
	 
	\begin{center}
	    { \LARGE{\bf Mass Gap without  Confinement}}
		
		\vspace{30pt}
		
		{}
		
		\vspace{25pt}
		
		{\large \bf Ant\'on F. Faedo,$^{1}$   David Mateos,$^{1,\,2}$   
David Pravos$^{1}$ and Javier G.~Subils$^{1}$}
		
\vspace{25pt}

{\normalsize  $^{1}$ Departament de F\'\i sica Qu\`antica i Astrof\'\i sica \& Institut de Ci\`encies del Cosmos (ICC),\\  Universitat de Barcelona, Mart\'\i\  i Franqu\`es 1, ES-08028, Barcelona, Spain.}\\
\vspace{15pt}
{ $^{2}$Instituci\'o Catalana de Recerca i Estudis Avan\c cats (ICREA), \\ Passeig Llu\'\i s Companys 23, ES-08010, Barcelona, Spain.}\\
		
		\vspace{70pt}

\begin{abstract}	

We revisit a one-parameter family of three-dimensional gauge theories with known supergravity duals. We show that three infrared behaviors are possible. For generic values of the parameter, the theories exhibit a mass gap but no confinement, meaning no linear quark-antiquark potential; for one limiting value of the parameter the theory flows to an infrared fixed point; and for another limiting value it exhibits both a mass gap and confinement. Theories close to these limiting values exhibit quasi-conformal and quasi-confining dynamics, respectively.  Eleven-dimensional supergravity provides a simple, geometric explanation of these features. 
\end{abstract}
	
\end{center}
	
\vspace{10pt}

\end{titlepage}

\baselineskip 5.5 mm


\tableofcontents

\section{Introduction and Summary} 
\label{intro}

\begin{figure}[t]
\begin{center}
\includegraphics[width=.95\textwidth]{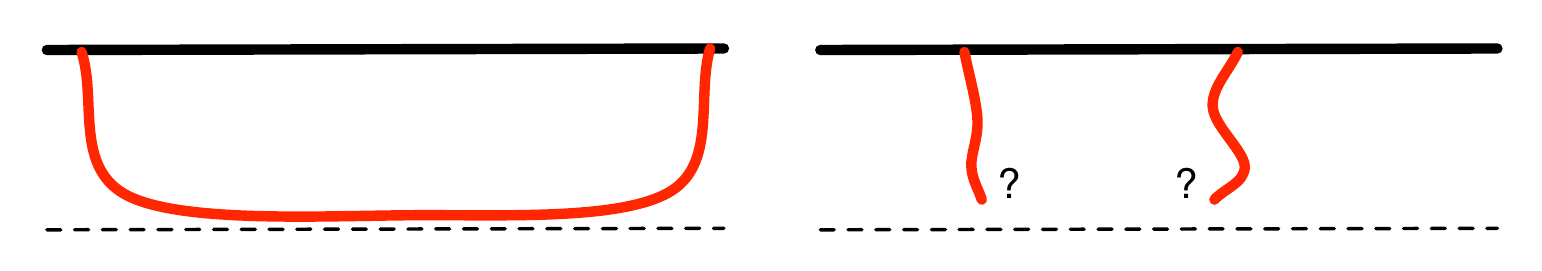} 
\caption{\small \small (Left) Connected string configuration (thick red curve) in the calculation of the quark-antiquark potential in string theory. The top, continuous, black, horizontal line represents the boundary on which the gauge theory resides. The bottom, dashed, black, horizontal line is the place where the geometry ends smoothly. (Right) Disconnected configuration that is not allowed due to charge conservation, since the endpoints of the strings have no place to end.}
\label{string}
\end{center}
\end{figure}

String theory duals of gauge theories that possess a mass gap are also expected to exhibit confinement in the sense of an asymptotically linear potential between an external quark-antiquark pair. Geometrically, the reason is simple, as illustrated in \Fig{string}. 
The mass gap arises because the geometry ends smoothly at  a non-zero value of the holographic coordinate. The linear potential comes from the fact that a string hanging from a well-separated quark-antiquark pair finds it  energetically advantageous to place most of its length near the regular end of the geometry, where it attains a constant, minimum energy  per unit length. A crucial ingredient in this argument is that the string configuration cannot consist of two disconnected pieces. The reason is charge conservation since, in a regular background, each piece would have no place to end. Put differently, an isolated quark or antiquark is not an allowed configuration.

\begin{figure}[t]
\begin{center}
\includegraphics[width=.95\textwidth]{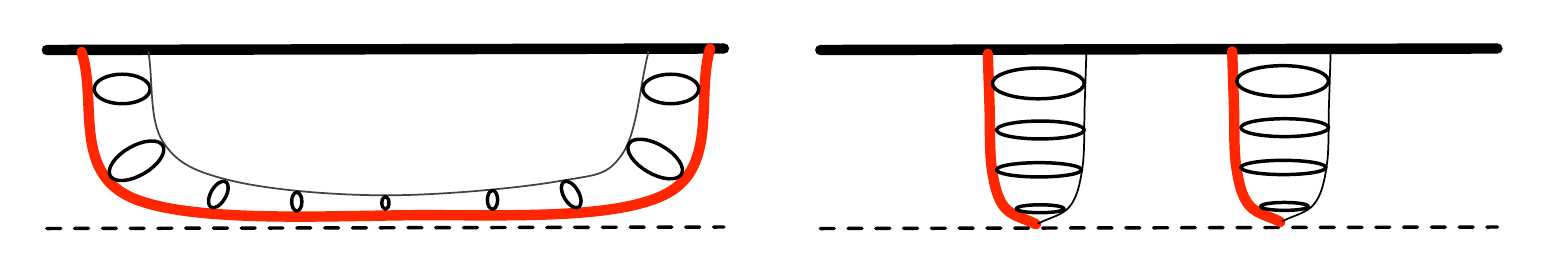} 
\caption{\small \small (Left) Connected membrane configuration in the calculation of the quark-antiquark potential in M-theory. The projection of the membrane onto the non-compact directions is represented by the thick, red curve. The M-theory circle at each point is represented next to it by the black circles. The top, continuous, black, horizontal line represents the boundary on which the gauge theory resides. The bottom, dashed, black, horizontal line is the place where the geometry ends smoothly. (Right) Disconnected membrane configuration  allowed by charge conservation, since the membrane closes off smoothly at the bottom of the geometry and hence it has a cigar-like topology with no boundary.}
\label{membrane}
\end{center}
\end{figure}

In this paper we will provide a counterexample to this expectation. The crucial point is that the gauge theories in question possess a regular supergravity description in eleven-dimensional M-theory but not in ten-dimensional string theory. Hence the existence of a mass gap or the presence of confinement can only  be reliably addressed in eleven dimensions. The eleven-dimensional geometries end smoothly at a non-zero value of the holographic coordinate, thus leading to a mass gap. However, no confinement arises. The reason is that, in M-theory, the quark-antiquark potential is calculated from the action of a membrane wrapped on the M-theory circle, which is just the uplift of the corresponding string calculation in ten dimensions. This is illustrated in \Fig{membrane}. 
In the geometries in question this circle shrinks smoothly to zero size in the infrared, leading to a cigar-like topology. A membrane wrapped on this cigar has no boundary and is thus compatible with charge conservation. In other words, an isolated quark or antiquark is an allowed configuration.  As a consequence, a configuration consisting of two cigar-like membranes hanging from the quark and the antiquark  at the boundary competes with the connected configuration and is in fact energetically preferred for a sufficiently large separation. It follows that  there is a phase transition from the connected to the disconnected configuration at a critical quark-antiquark separation that cuts-off the linear growth of the potential. 

The geometries that realize the physics above consist of a one-parameter family of supergravity solutions dual to a one-parameter family of three-dimensional gauge theories. We emphasize that these supergravity solutions themselves are not new \cite{Cvetic2}, but we present them in what we hope is a user-friendly, comprehensive treatment. Throughout the paper we will find it useful to switch back and forth between the descriptions in ten and eleven dimensions. All solutions preserve two supercharges, corresponding to $\mathcal{N}=1$ supersymmetry in three dimensions. Because of the small amount of supersymmetry it is difficult to determine the precise details of the  dual gauge theories. Nevertheless, they are presumably  quiver-like, super Yang--Mills (SYM)  gauge theories with a product gauge group of the form 
$\mbox{U($N+M$)} \times \mbox{U($N$)}$ and possible additional Chern--Simons--Matter terms (CSM) \cite{Loewy,Hashimoto}. For brevity, we will refer to these as SYM-CSM theories. 

Each of the eleven-dimensional solutions is based on an eight-dimensional transverse geometry of Spin(7) holonomy,\footnote{Except for the one labelled 
$\mathbb{B}_8^{\rm{conf}}$, see below.\label{below}} as we will review in Sec.~\ref{sing}. Given the transverse geometry, the two additional ingredients needed to obtain the corresponding eleven-dimensional solution are appropriate fluxes through the transverse geometry and a warp factor. If the ranks of the two gauge groups are the same, i.e.~if $M=0$, then the resulting warp factor is singular, and so is the eleven-dimensional solution, as we explain in Sec.~\ref{sing}. In order to have $M\neq 0$ one must add fractional branes to the system, as we review in Sec.~\ref{adding}, which results in additional fluxes. Under these circumstances it is then possible to construct completely regular eleven-dimensional solutions, as we show in Sec.~\ref{regular}. 

\begin{figure}[t]
\centering
\begin{tikzpicture}[scale=3.3,very thick,decoration={markings,mark=at position .5 with {\arrow{stealth}}}]

	\node[above] at (0,0) {SYM-CSM $|$ D2};
	\node[below] at (0,-2.2) {Mass gap $|$ $\mathbb{R}^4\times {\rm S}^4$};
	\node[left] at (-1,-1) {OP $|$ CFT};
	\node[right] at (2,-2) {Confinement $|$ $\mathbb{R}^3\times{\rm S}^1\times {\rm S}^4$};
	
	\draw [black, ultra thick] (0,0) circle [radius=0.015];
	\draw [black, ultra thick] (-1,-1) circle [radius=0.015];

	\node at (-.6,-1) {$\mathbb{B}_8^+$};
	\node at (.63,-1) {$\mathbb{B}_8^-$};

	\draw[postaction={decorate},very thick] (0,0) --  (0,-2) node[left,midway]{$\mathbb{B}_8$};
	\draw[postaction={decorate},ultra thick] (0,0) -- (-1,-1) node[left,midway]{$\mathbb{B}_8^\infty$\,};
	\draw[postaction={decorate},ultra thick] (0,0) -- (2,-2) node[right,midway]{\, $\mathbb{B}_8^{\rm{conf}}$};
	\draw[postaction={decorate},ultra thick] (-1,-1) -- (-1,-2) node[left,midway]{$\mathbb{B}_8^{\textrm{\tiny OP}}$\,\,};

        \draw[postaction={decorate},thick, red] (0,0) .. controls (-.9,-.9) and (-1,-1) .. (-0.98,-2);
	\draw[postaction={decorate},thick, red] (0,0) .. controls (-.5,-.5) and (-.5,-1.5) .. (-.5,-2);
	\draw[postaction={decorate},thick, blue] (0,0) .. controls (.5,-.5) and (.5,-1.5) .. (.5,-2);
	\draw[postaction={decorate},thick, blue] (0,0) .. controls (.9,-.9) and (.95,-1.05) .. (1,-2);
	\draw[postaction={decorate},thick, blue] (0,0) .. controls (.9,-.9) and (1.5,-1.6) .. (1.5,-2);
	\draw[postaction={decorate},thick, blue] (0,0) .. controls (.95,-.95) and (1.8,-1.8) .. (1.9,-2);
	
	\draw[|-|] (-1,-2) -- (0,-2);
	\draw[-stealth] (0,-2) -- (2,-2);
	
	\node[left=5] at (-1,-2) {$y_0$};
	\node[below=5] at (-1,-2) {$-1$};
	\node[below=5] at (0,-2) {$1$};
	\node[below=5] at (2,-2) {$\infty$};
	
\end{tikzpicture}
\caption{\small Pictorial representation of the different solutions (see main text).
}\label{fig:triangle}
\end{figure}
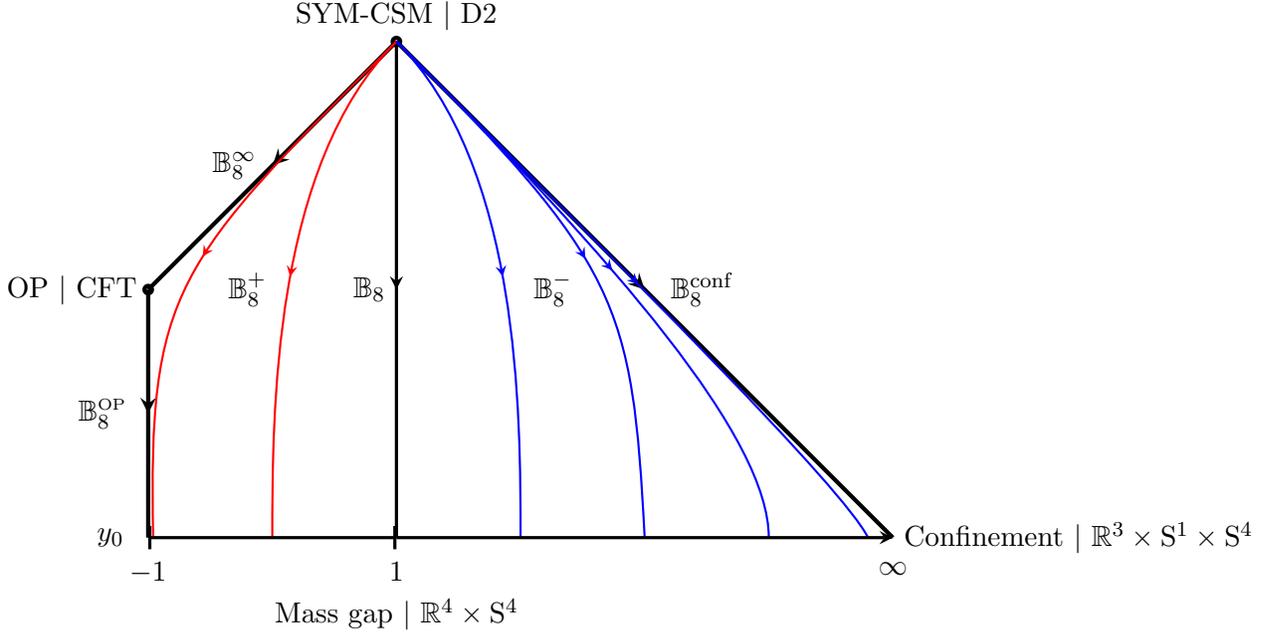

The set of solutions is pictorially summarized in \Fig{fig:triangle}. Each curve or straight line running downwards represents an eleven-dimensional solution labelled by its corresponding eight-dimensional  transverse  geometry. The arrows indicate the direction of the Renormalization Group (RG) flow from the ultraviolet (UV) to the infrared (IR). In the UV all solutions$^{\ref{below}}$
 are dual to a SYM-CSM theory, as described above. Asymptotically, the corresponding ten-dimensional geometries are those of D2-branes placed at a cone over $\mathbb{CP}^3$, hence the label at the top vertex of the figure. These geometries are accompanied by two types of fluxes. First,$^{\ref{below}}$ an ABJM-like \cite{Aharony} flux of the Ramond-Ramond (RR) two-form proportional to the K\"ahler form on $\mathbb{CP}^3$. As in ABJM, we expect this to indicate the presence of CSM interactions and to determine the CS level. Second, fluxes associated to the presence of fractional branes that render the IR metrics regular and shift the rank of one of the two gauge groups.  

The family of supergravity solutions is parametrized by a constant that we call $y_0$ and that takes values between $-1$ and $\infty$. We expect this parameter to be related to the difference between the  couplings of the two factors in the gauge group, as we will comment further in \Sec{regular}. 
All the transverse eight-dimensional geometries can be foliated by squashed seven-spheres viewed as an ${\rm S}^1$ fibration over an ${\rm S}^2$ base that is itself fibered over an ${\rm S}^4$. Thus one can also view these geometries as a squashed ${\rm S}^3$ fibered over ${\rm S}^4$, or as an ${\rm S}^1$ fibered over a squashed 
$\mathbb{CP}^3$. 
Following the original references, we refer to solutions with $y_0 \in (-1,1)$ as the $\mathbb{B}_8^+$ family \cite{Cvetic2}, to the solution with $y_0=1$ as $\mathbb{B}_8$ \cite{Cvetic3}, and to solutions with 
$y_0 \in (1,\infty)$ as the $\mathbb{B}_8^-$ family \cite{Cvetic2}. Despite this technical distinction, the physics is continuous as a function of $y_0$ for $y_0 \in (-1,\infty)$. In this set of solutions the ${\rm S}^3$ shrinks smoothly to zero size in the IR, whereas the size of the ${\rm S}^4$ remains non-zero. The IR transverse   geometry is thus $\mathbb{R}^4\times {\rm S}^4$. We will see that this leads to a mass gap but no confinement. 

For $y_0=-1$ the IR physics is radically different. The transverse geometry was found and dubbed $\mathbb{B}_8^\infty$ in \cite{Hashimoto}.  In this case the entire ${\rm S}^7$  shrinks to zero size in the IR, but once the warp factor is taken into account the resulting geometry is $\mbox{AdS}_4$ times a squashed seven-sphere of non-zero size. This fixed point is dual to the so-called Ooguri--Park (OP) conformal field theory (CFT) \cite{Ooguri}, which is an $\mathcal{N}=1$ deformation of the  ABJM  theory \cite{Aharony}. 
Remarkably, the OP fixed point admits a relevant deformation that drives it to an IR theory with a mass gap but no confinement. The eleven-dimensional solution  describing this flow is based on a transverse geometry found in \cite{Bryant, Gibbons}  that we call $\mathbb{B}_8^{\textrm{\tiny OP}}$. As we will explain in Sec.~\ref{quasiconformal}, solutions with $y_0$ close to $-1$ describe RG flows that approach the concatenation of the $\mathbb{B}_8^\infty$ and the $\mathbb{B}_8^{\textrm{\tiny OP}}$ flows. These solutions exhibit ``walking'' or quasi-conformal behavior in a certain range of energies, as we will confirm in Sec.~\ref{potential} with a calculation of the potential between an external quark-antiquark pair. 

If the RR two-form is set to zero, one obtains a solution based on an internal geometry found in \cite{Bryant, Gibbons} that we call $\mathbb{B}_8^{\rm{conf}}$. The eleven-dimensional solution flows to an IR theory that exhibits both a mass gap and confinement. The geometric reason is that, in this case, the S$^1$ is trivially fibered over the rest of the  geometry and it remains non-contractible along the entire flow; in particular, the IR transverse geometry is $\mathbb{R}^3\times{\rm S}^1\times {\rm S}^4$.  This implies that a membrane wrapped on this S$^1$ cannot end anywhere in the bulk since it would have a cylinder-like geometry and hence a boundary, which is not allowed by charge conservation. On the gauge theory side, the existence of confinement  seems to be a consequence of the absence of CSM interactions. We will confirm the presence of an asymptotic linear potential for an external quark-antiquark pair in Sec.~\ref{potential}. We will also show in Sec.~\ref{quasiconfining} that the $\mathbb{B}_8^{\rm{conf}}$ can be obtained as the $y_0 \to \infty$ limit of the $\mathbb{B}_8^-$ solutions.

\section{Preliminaries}
\label{preli}

As explained in Sec.~\ref{intro}, the type IIA solutions of interest in this paper  describe RG flows that start from a D2-brane-like asymptotic geometry in the UV. Some of these flows end at AdS$_4$ geometries in the IR. Moreover, throughout the paper we will be switching between the type IIA description of these solutions in ten dimensions and their M-theory description in eleven dimensions. We therefore begin by reviewing a few facts about the simplest examples of these kinds of  solutions. 

The metric and dilaton of the type IIA solution describing $N$ D2-branes at the tip of a manifold of G$_2$  holonomy take the form
 \begin{eqnarray}
 \label{D2brane}
  \dd s_{\rm st}^2&=&h^{-\frac12}\,\dd x_{1,2}^2+h^{\frac12}\, 
  \dd s^2 (\mbox{M}_7) \,,\nonumber
  \\[2mm]
  e^\Phi&=&h^{\frac14}  \,.
  \end{eqnarray}
The requirement of G$_2$  holonomy guarantees that the solution preserves $\mathcal{N}=1$ supersymmetry in three dimensions, i.e.~two supercharges. If  the transverse seven-dimensional space is a cone then the base must be a nearly-K\"ahler six-dimensional manifold $\left({\rm NK}_6\right)$:
\begin{equation}
\label{cone}
\dd s^2 (\mbox{M}_7)\,=\,\dd r^2+r^2\,\dd s^2\left({\rm NK}_6\right)\,.
\end{equation}
In terms of the radial coordinate on the cone, the warp factor in \eqq{D2brane} behaves as 
\be
h\sim \frac{N}{r^{5}} \,.
\ee
In the particular case that the NK manifold is S$^6$ the supersymmetry is enhanced to $\mathcal{N}=8$ (i.e.~sixteen supercharges) and the gauge theory dual is maximally supersymmetric YM in three dimensions \cite{Itzhaki}. In the present  paper the NK manifold of interest is $\mathbb{CP}^3$, in which case the gauge theory dual is expected to be a quiver YM-type theory with gauge group  $\mbox{U(N)} \times \mbox{U(N)}$ \cite{Loewy}. 

The uplift of the D2-brane solution to eleven dimensions is straightforwardly obtained via the usual ansatz
\begin{equation}
\dd s_{11}^2\,=\,e^{-\frac23\Phi}\,\dd s_{\rm s}^2+e^{\frac43\Phi}\,\ell_p^2 \,
\left(\dd\psi+C_1\right)^2\,,
\end{equation}
where $\psi$ parameterizes the M-theory  circle, $\ell_p$ is the eleven-dimensional Planck length and $C_1$ is the RR one-form potential of type IIA. Since the D2-brane solution has $C_1=0$, the result is the M2-brane-type metric 
\begin{equation}
\label{M2metric}
\dd s_{11}^2\,=\,H^{-2/3}\,\dd x_{1,2}^2+H^{1/3}\,\dd s_8^2\,,
\end{equation}
with
\begin{equation}
H\,=\,e^{\Phi}\,h^{3/4}\,=\,h\sac
\dd s_8^2\,=\,\dd r^2+r^2 \dd s^2\left({\rm NK}_6\right)+\ell_p^2\, \dd  \psi^2\,.
\end{equation}
We see that the fact that in the D2-brane solution the RR two-form vanishes implies that the M-theory circle is trivially fibered over the rest of the directions, and that it has constant size in the eight-dimensional transverse metric.

Let us contrast this with the uplift of the ${\rm AdS}_4 \times \mathbb{CP}^3$ solution of type IIA supergravity, whose gauge theory dual is the ABJM CSM theory  \cite{Aharony}. In this case we view $\mathbb{CP}^3$ as a  K\"ahler manifold instead of as a NK manifold (see below for more details on this distinction). The RR two-form is proportional to the K\"ahler form, the dilaton is constant and the warp factor scales as $h \sim r^{-4}$, which results in the uplift 
\begin{equation}
H\,=\,h^{3/4} \sim r^{-3}\sac
\dd s_8^2\,=\,\dd r^2+r^2\dd s^2 (\mathbb{CP}^3)+r^2\left(\dd\psi+C_1\right)^2 \,.
\end{equation}
It can be checked that the metric in eleven dimensions contains again an 
${\rm AdS}_4$ factor. We see how the M-theory circle is non-trivially fibered over, and that its size grows as that of the other directions. 

The solutions of interest in the rest of the paper are based on the eight-dimensional Spin(7)-holonomy metrics of  \cite{Cvetic2}, which combine ingredients of the two cases above. On the one hand, the M-theory circle is non-trivially fibered over the rest of the coordinates. On the other hand, its size does not grow asymptotically with the other directions but approaches a constant. In the type IIA description this means that the metric and the dilaton behave asymptotically as in the D2-brane solution \eqq{D2brane} but, unlike in the pure D2-brane solution, the RR two-form does not vanish. For these reasons we expect the dual gauge theory in the UV to be a quiver  SYM theory with gauge group 
$\mbox{U(N)} \times \mbox{U(N)}$ and additional CSM terms. 

\section{Geometry of $\mathbb{CP}^3$}

Since this manifold will play a crucial role in our solutions, we will discuss some of its properties in this section. For our purposes, a useful way to describe $\mathbb{CP}^3$ is as the twistor space over the four sphere, or in other words, as an S$^2$ fibration over S$^4$:
\begin{equation}
\begin{array}{rcl}
{\rm S}^2&\hookrightarrow&\mathbb{CP}^3={\rm Tw}\left({\rm S}^4\right)\\[-2mm]
&&\,\,\,\downarrow\\[-2mm]
&&\,\,\,{\rm S}^4
\end{array}
\end{equation}
This non-trivial fibration allows us to consider deformations in which we squash the fiber with respect to the base. A convenient set of coordinates was introduced in \cite{Conde}, and we follow their notation with slight differences. In terms of the vielbeins $E^i$ to be defined below, the metric can be written as
\begin{equation}
\label{CP3metric}
\dd s_6^2\,=\,\alpha^2\,\left[\left(E^1\right)^2+\left(E^2\right)^2\right]+\dd \Omega_4^2\,,
\end{equation}
where the metric in brackets is the metric on a round S$^2$ and $E^1, E^2$ describe the non-trivial fibration (i.e.~they contain coordinates of S$^4$). We have included the squashing parameter $\alpha$ that controls the size of the fiber with respect to the base. There are two special values of this parameter for which the metric becomes Einstein: $\alpha^2=1$ and $\alpha^2=1/2$. If $\alpha^2=1$ we recover the unsquashed $\mathbb{CP}^3$ with the Fubini--Study metric, which is K\"ahler. This is the metric appearing in the ABJM construction \cite{Aharony}. If  $\alpha^2=1/2$, the metric admits instead a nearly K\"ahler structure, as in \eqq{cone}. This is the metric that was used in the construction of \cite{us}, where unquenched flavor was added to three-dimensional SYM. 

There is another special point, $\alpha^2=1/5$, for which the metric, despite it not being Einstein, supports a minimally supersymmetric AdS solution. Its uplift to M-theory corresponds to the squashed seven-sphere, and the dual gauge theory is an $\mathcal{N}=1$ deformation of ABJM \cite{Ooguri}.

Two more important facts about this geometry are the following. First, the isometry group of the metric (\ref{CP3metric}) is generically ${\rm Sp}(2)\sim {\rm SO}(5)$, which is enhanced to ${\rm SU}(4)\sim {\rm SO}(6)$ at the special point $\alpha^2=1$. This means that it will  be convenient to describe 
$\mathbb{CP}^3$ as the coset ${\rm Sp}(2)/{\rm U}(2)$, since we are interested in solutions preserving these isometries. Second, the non-vanishing Betti numbers are 
\be
\label{betti}
b_0=b_2=b_4=b_6=1 \,, 
\ee
meaning that $\cpthree$ possesses non-trivial two- and four-cycles. 

In constructing the solutions, an important ingredient is the set of Sp(2)-left-invariant forms on the coset ${\rm Sp}(2)/ {\rm U}(2)$. We will be using the coordinate system in \cite{Conde} to facilitate the comparison, although a coordinate system is not indispensable. Given the SU(2)-left-invariant forms 
$\omega^i$, verifying 
\be
\dd \omega^i=\frac12 \epsilon_{ijk}\omega^j\wedge\omega^k \,, 
\ee
the metric of the four-sphere can be written as
\begin{equation}
\dd \Omega_4^2\,=\,\frac{4}{\left(1+\xi^2\right)^2}\left[\dd\xi^2+\frac{\xi^2}{4}\omega^i\omega^i\right]\,,
\end{equation}
with $\xi$ a non-compact coordinate. If the S$^2$ fiber is parameterized by the usual angles $\theta$ and $\varphi$, then the non-trivial fibration is described by the vielbeins 
\begin{eqnarray}
\label{introduced}
E^1&=&\dd \theta+\frac{\xi^2}{1+\xi^2}\left(\sin\varphi\,\omega^1-\cos\varphi\,\omega^2\right)\,,\nonumber\\[2mm]
E^2&=&\sin\theta\left(\dd\varphi-\frac{\xi^2}{1+\xi^2}\omega^3\right)+\frac{\xi^2}{1+\xi^2}\cos\theta\left(\cos\varphi\,\omega^1+\sin\varphi\,\omega^2\right)\,.
\end{eqnarray}
For our purposes, it is convenient to consider a rotated version of the vielbeins on the four-sphere that read\footnote{With respect to \cite{Conde}, we are taking $\mathcal{S}^\xi_{\textrm{\tiny there}}=\mathcal{S}^4_{\textrm{\tiny here}}$ and $\mathcal{S}^3_{\textrm{\tiny there}}=-\mathcal{S}^3_{\textrm{\tiny here}}$.}
\begin{eqnarray}
\mathcal{S}^1&=&\frac{\xi}{1+\xi^2}\left[\sin\varphi\,\omega^1-\cos\varphi\,\omega^2\right]\,,\nonumber\\
\mathcal{S}^2&=&\frac{\xi}{1+\xi^2}\left[\sin\theta\,\omega^3-\cos\theta\left(\cos\varphi\,\omega^1+\sin\varphi\,\omega^2\right)\right]\,,\nonumber\\
\mathcal{S}^3&=&\frac{\xi}{1+\xi^2}\left[\cos\theta\,\omega^3+\sin\theta\left(\cos\varphi\,\omega^1+\sin\varphi\,\omega^2\right)\right]\,,\nonumber\\
\mathcal{S}^4&=&\frac{2}{1+\xi^2}\,\dd\xi\,.
\end{eqnarray}
Despite the fact that these forms depend on the S$^2$ angles, it is easily checked that 
\be
\mathcal{S}^n\mathcal{S}^n=\dd \Omega_4^2 \,.
\ee
In terms of these, the left-invariant two-forms on the coset are
\begin{equation}
\label{2forms}
X_2\,=\,E^1\wedge E^2\,,\qquad\qquad\qquad J_2\,=\,\mathcal{S}^1\wedge\mathcal{S}^2+\mathcal{S}^3\wedge\mathcal{S}^4\,.
\end{equation}
Similarly, the globally defined, left-invariant  three-forms are:
\begin{eqnarray}
\label{3forms}
X_3&=&E^1\wedge\left(\mathcal{S}^1\wedge\mathcal{S}^3-\mathcal{S}^2\wedge\mathcal{S}^4\right)-E^2\wedge\left(\mathcal{S}^1\wedge\mathcal{S}^4+\mathcal{S}^2\wedge\mathcal{S}^3\right)\,,
\nonumber\\[2mm]
J_3&=&-E^1\wedge\left(\mathcal{S}^1\wedge\mathcal{S}^4+\mathcal{S}^2\wedge\mathcal{S}^3\right)-E^2\wedge\left(\mathcal{S}^1\wedge\mathcal{S}^3-\mathcal{S}^2\wedge\mathcal{S}^4\right)\,.
\end{eqnarray}
Finally, the invariant four-forms are the wedges of the two-forms 
\be
\label{4forms}
X_2\wedge J_2 \sac J_2\wedge J_2=2\epsilon_{(4)} \,, 
\ee
where $\epsilon_{(n)}$ denotes the volume-form of the $n$-sphere. Left-invariance ensures that this system of forms closes under exterior differentiation and Hodge duality. In particular we have
\begin{eqnarray}\label{extder}
\dd X_2&=&\dd J_2\,=\,X_3\,,\qquad\qquad\dd J_3\,=\,2\left(X_2\wedge J_2+J_2\wedge J_2\right)\,,\nonumber\\[2mm]
*X_2&=&\frac{1}{2\alpha^2} J_2\wedge J_2\,,\qquad\qquad*J_2\,=\,\alpha^2 X_2\wedge J_2\,,\qquad\qquad*X_3\,=\,-J_3\,.
\end{eqnarray}
From these forms it is easy to construct both the K\"ahler and the nearly K\"ahler structures on  $\mathbb{CP}^3$. When the squashing in (\ref{CP3metric}) is fixed to $\alpha^2=1$ the metric admits a K\"ahler structure, whose K\"ahler form is
\begin{equation}
\label{Kahlerform}
J_{\textrm {\tiny K}}\,=\,X_2-J_2\,,
\end{equation} 
which is closed by virtue of (\ref{extder}). If instead  the squashing is 
$\alpha^2=1/2$, the almost-complex structure associated to the NK structure reads
\begin{eqnarray}
J_{\textrm {\tiny NK}}&=&\frac12X_2+J_2 \,.
\end{eqnarray}
This shows that the set of Sp(2)-invariant forms is general enough for our purposes.  In the following we will use them to construct 
solutions of type IIA supergravity with $\mathbb{CP}^3$ as their internal geometry. 

\section{Singular Flows}
\label{sing}

As explained in Sec.~\ref{intro}, in this section we will construct type IIA solutions describing RG flows from a D2-brane-like asymptotic geometry in the UV to a singular geometry in the IR. The uplifts of these solutions to M-theory are also singular in the IR. In the following sections we will modify these solutions in such a way that their eleven-dimensional description is completely regular. 

The transverse, seven-dimensional geometries that we will employ  are the dimensional reduction of the eight-dimensional metrics found in \cite{Cvetic2}, with which we will make contact below. Despite the fact that our UV asymptotic geometries are different from those in \cite{Conde}, which focused on AdS$_4$ solutions, the metrics that we are interested in fall within the ansatz studied in \cite{Conde}, which we therefore follow.

The ten-dimensional string-frame metric and dilaton take the form
\begin{eqnarray}
\label{runningD2}
\dd s_{\rm st}^2 &=&h^{-\frac12}\,\dd x_{1,2}^2+h^{\frac12}\, \dd s^2_7
\,,\nonumber\\[2mm]
e^\Phi&=&h^{\frac14} \, e^\Lambda \,,
\end{eqnarray}
with the transverse geometry given by 
\begin{equation}\label{SquashedCP3}
\dd s_7^2\,=\,\dd r^2+e^{2f}\,\dd\Omega_4^2+e^{2g}\,\left[\left(E^1\right)^2+\left(E^2\right)^2\right] \,.
\end{equation}
The warp factor $h$, the squashing functions $f,g$ and the dilaton function $\Lambda$ depend only on the radial coordinate $r$. Note that $r$, $e^{f}$ and  $e^{g}$ have dimensions of length, whereas $h$ is dimensionless. The D2-brane solution \eqq{D2brane}, to which our more general solutions will asymptote, is recovered setting
\be
h\sim \frac{N}{r^{5}}\,, \qquad\qquad e^{2f}=\frac{1}{2} r^2\,, \qquad\qquad e^{2g}=\frac{1}{4} r^2\,, \qquad\qquad
e^\Lambda=1 \,.
\ee

The metric and dilaton \eqq{runningD2} will be supported by the fluxes
\begin{eqnarray}
F_2&=&Q_k\,J_{\rm K} \,,\label{F2flux} \\[2mm]
F_4&=&\dd^3x\wedge\dd\left(h^{-1}\,e^{-\Lambda}\right)\,, \label{F4flux}
\end{eqnarray}
where we recall that $J_{\rm K}$ is the K\"ahler form of  $\cpthree$ given in \eqq{Kahlerform}. The fact that $F_4$  does not involve any new functions beyond those appearing in the metric and the dilaton is a reflection of supersymmetry. Closure of $F_2$ implies that $Q_k$ is a constant. 

The first-order BPS equations ensuring  $\mathcal{N}=1$ supersymmetry  follow  from the results in \cite{Conde} and read
\begin{eqnarray}
\label{BPSsystem}
\Lambda'&=&2Q_k\,e^{\Lambda-2f}-Q_k\,e^{\Lambda-2g}\,,\nonumber\\[2mm]
f'&=&\frac{Q_k}{2}\,e^{\Lambda-2f}-\frac{Q_k}{2}\,e^{\Lambda-2g}+e^{-2f+g}\,,\\[2mm]
g'&=&Q_k\,e^{\Lambda-2f}+e^{-g}-e^{-2f+g}\,.\nonumber
\end{eqnarray}
The  warp factor can be expressed in terms of the other functions as \cite{Conde}
\begin{equation}
h\,=\,e^{-\Lambda}\left[h_0-Q_c\int^re^{2\Lambda(z)-4f(z)-2g(z)}\dd z\right]\,.
\end{equation}
The constant $Q_c$ is related to the number of D2-branes, as we will see below, and has dimensions of (length)$^{5}$. The integrand has dimensions of (length)$^{-6}$ and $h$ is dimensionless. 
The integration constant $h_0$ can be shifted by changing the lower limit of the integral, so we will henceforth set $h_0=0$ without loss of generality. 

The usual quantization condition for the RR fluxes takes the form
\begin{equation}\label{chargenofrac}
\int_{\Sigma_{8-p}}F_{8-p}\,=\,2\kappa_{10}^2T_{{\rm D}p}\,N_{p}\,,
\end{equation}
where $\Sigma_{8-p}$ is an appropriate cycle. In the case $p=6$ this cycle is a $\mathbb{CP}^1\subset\mathbb{CP}^3$ given by constant coordinates on the $\rm{S}^4$ and we get 
\begin{equation}
\label{Qk}
Q_k\,=\,\frac{\ell_s g_s}{2}\,k\,.
\end{equation}
As in \cite{Aharony,Ooguri}, we expect $k$ to be the CS level of the dual gauge theory. In the case $p=2$ the cycle is the entire $\mathbb{CP}^3$ and we find
\begin{equation}
Q_c\,=\,3\pi^2\ell_s^5\,g_s\,N\,,
\label{Qc}
\end{equation}
where $N$ is the number of D2-branes and the rank of the field theory gauge group.

In order to make contact with \cite{Cvetic2}, let us uplift our ansatz to eleven dimensions.  The K\"ahler form (\ref{Kahlerform}) can be written as 
$J_{\textrm {\tiny K}} = \dd C_1$ with the potential 
\begin{equation}
C_1\,=\,-\left(\cos\theta\,\dd \varphi-\xi\,\mathcal{S}^3\right) \,.
\end{equation}
This means that in terms of the vielbein
\begin{equation}
E^3\,=\,\dd\psi-\cos\theta\,\dd \varphi+\xi\,\mathcal{S}^3 \,,
\end{equation}
with $\psi\in\left[0,\frac{4\pi}{k}\right)$,  the eleven-dimensional metric can be written in the M2-brane form \eqq{M2metric} with 
\begin{equation}
\dd s_8^2\,=\,e^{-\Lambda}\,\Bigg[ \dd r^2+e^{2f}\,\dd\Omega_4^2+e^{2g}\,\left[\left(E^1\right)^2+\left(E^2\right)^2\right] \Bigg]+e^\Lambda\,Q_k^2\,\left(E^3\right)^2
\end{equation}
and
\be
\label{HH}
H\,=\, h \, e^\Lambda \,.
\ee
Comparing with the ansatz in \cite{Cvetic2} we see that the functions $a, b, c$ used there are related to ours through 
\begin{equation}
\label{abc}
a^2\,=\,e^{2g-\Lambda}\,,\qquad\qquad b^2\,=\,Q_k^2\,e^\Lambda\,,\qquad\qquad c^2\,=\,e^{2f-\Lambda}\,,
\end{equation}
and that their equations for special holonomy are equivalent to the BPS system (\ref{BPSsystem}). To be precise, in order to recover the results of \cite{Cvetic2} we should set $Q_k=-1$. Presumably, the sign is  a choice of radial coordinate, and  absolute values different from one  correspond to orbifolds of the construction in \cite{Cvetic2}. In most of our paper we will focus on the case with negative $Q_k$.  We conclude that all the solutions of \cite{Cvetic2} are also solutions of our equations and that we have directly constructed their ten-dimensional description. 

Before we describe the family of metrics that will be our main interest, let us point out two particularly simple solutions that are dual to superconformal theories. If $Q_k>0$ there is the ABJM fixed point \cite{Aharony} described by
\begin{equation}
e^{2f}\,=\,r^2\,,\qquad\qquad e^{2g}\,=\,r^2\,,\qquad\qquad e^{\Lambda}\,=\,\frac{r}{Q_k}\,.
\end{equation}
The supersymmetry of this solution is enhanced generically to $\mathcal{N}=6$. 
If  $Q_k<0$ there is the OP fixed point \cite{Ooguri} at
\begin{equation}
\label{OPsolution}
e^{2f}\,=\,\frac95\,r^2\,,\qquad\qquad e^{2g}\,=\,\frac{9}{25}\,r^2\,,\qquad\qquad e^{\Lambda}\,=\,\frac{3r}{5|Q_k|} \,.
\end{equation}
The theory dual to the OP fixed point is an $\mathcal{N}=1$ deformation of the ABJM model. It is easy to show that in both cases the coefficients of the three vielbeins $E^i$ in the uplift coincide and that altogether they parameterize a round three-sphere fibered over ${\rm S}^4$. Thus in the ABJM case the eleven-dimensional geometry is 
$\mbox{AdS}_4$ times a round $\mbox{S}^7$ (orbifolded by $\mbox{Z}_k$), whereas in the OP case the sphere is not round but squashed. For this reason, upon reduction to ten dimensions the internal metric involves the unsquashed 
$\cpthree$ in the ABJM case, corresponding to $\alpha^2=1$ in \eqq{CP3metric}, whereas for OP it involves the squashed $\cpthree$ with $\alpha^2=1/5$.
 
We now proceed to describe the family of metrics that will be our main interest. 
The general solution to our BPS system can be found using the tricks developed in \cite{Cvetic2,Cvetic3}. First we define the master function 
\be
P(r')=e^{2f-\Lambda} \,,
\ee
which verifies the third-order equation
\begin{equation}
\left(P'-Q_k\right)T=P\,T'\,,
\end{equation}
where
\begin{equation}
T\,=\,2\,P\,W'+\left(P'+3Q_k\right)W\,,\qquad\qquad\qquad W\,=\,P'+Q_k\,.
\end{equation}
Note that $P$ has dimensions of (length)$^2$. 
We now change to a radial coordinate $\varrho$ and a function $\gamma(\varrho)$ defined by the conditions
\begin{equation}\label{gamma}
P'\,=\,\varrho\,,\qquad\qquad\qquad P''\,=\,-\frac{\gamma}{P}\,.
\end{equation}
With this we reduce the system to the first-order equation
\begin{equation}
\gamma\left(2\frac{\dd \gamma}{\dd \varrho}  -6Q_k\right)\,=\,\left(\varrho+3Q_k\right)\left(\varrho^2-Q_k^2\right)\,.
\end{equation}
The final change of variables to a new radial coordinate $y$ and a new function $v(y)$ defined through 
\begin{equation}\label{ycoordinate}
y\,=\,\frac{2\left(Q_k^2-\gamma+Q_k\varrho\right)}{\left(Q_k+\varrho\right)^2}\,,\qquad\qquad\qquad \varrho\,=\,-Q_k\left(v+1\right) \,,
\end{equation}
linearizes the equation to
\begin{equation}
2\left(1-y^2\right)\frac{\dd v}{\dd y}\,=\,y\,v+2\,.
\label{veq}
\end{equation}
Note that $y$ and $v$ are both dimensionless. This equation can be solved in terms of generalized hypergeometric functions, as we will explain below. 
Going back to  (\ref{gamma}), we see that the master function satisfies the equation
\begin{equation}
\label{Peq}
\frac{1}{P}\,\frac{\dd P}{\dd y}\,=\,\frac{v+1}{v\left(1-y^2\right)}\,.
\end{equation}
The rest of the functions are determined in terms of $P$ as
\begin{equation}
\label{looking2}
e^g\,=\,\frac{2P\left(2-v\right)}{Q_k\left(1+y\right)v^2}\,,\qquad\qquad\qquad e^{\Lambda}\,=\,\frac{4P\left(v-2\right)}{Q_k^2v^3\left(1+y\right)}\,.
\end{equation}
Following the chain of definitions, we see that the $r$ and $y$ coordinates are related through 
\begin{equation}\label{ry}
\dd r\,=\,-\frac{P}{Q_k\,v\,\left(1-y^2\right)}\dd y\,,
\end{equation}
and hence that the eight-dimensional, Spin(7)-holonomy metric takes the form 
\vspace{2mm}
\begin{equation}
\label{eight}
\dd s_8^2\,=\,\frac{v\,P\,\left(1+y\right)\dd y^2}{4\left(v-2\right)\left(1-y^2\right)^2}+P\,\dd \Omega_4^2+\frac{P\left(v-2\right)}{v\left(1+y\right)}\left[\left(E^1\right)^2+\left(E^2\right)^2\right]+\frac{4P\left(v-2\right)}{v^3\left(1+y\right)}\left(E^3\right)^2 \,.
\end{equation}

The general solution of \eqq{veq} splits into several families depending on the initial conditions for the ``flow'' $v(y)$. We are particularly interested in the  families denoted $\mathbb{B}_{8}^+$ and $\mathbb{B}_{8}^-$ in \cite{Cvetic2, Cvetic3}. Both are characterized by the fact that there is a value of the radial coordinate, $y=y_0$, such that  
\be
\label{condition}
v\left(y_0\right)=2 \,. 
\ee 
At this point  the three-sphere parametrized by $E^i$ in \eqq{eight} shrinks smoothly to zero size, whereas the size of the four-sphere remains finite. 
We will see that, in both families, the IR region lies at $y\to y_0$ and the UV at $y \to 1$. Since the allowed values of $y_0$ are different in each case,  we will consider each family separately. 

Let us finally mention that, aside from the solutions of $\mathbb{B}_8$-type that we are discussing in this paper, the only other solution to the system (\ref{BPSsystem}) that provides a physically acceptable  metric is the one dubbed $\mathbb{A}_8$ in \cite{Cvetic2}. This geometry, which exists  for $Q_k>0$, describes a flow that starts at the same UV theory as the $\mathbb{B}_8$ metrics, but that in the IR flows to the ABJM fixed point. The rest of the solutions of (\ref{BPSsystem}) are either singular and/or produce signature changes in the metric.

\subsection{$\mathbb{B}_{8}^+$ family}

In this case the range of the radial coordinate is  
\be
-1 \leq y_0 \leq y < 1\,,
\ee
and the solutions of \eqq{veq} and \eqq{Peq} are
\begin{eqnarray}
v&=&v^+ (y)\,=\,\frac{1}{\left(1-y^2\right)^{1/4}}\,\left(v_0^+ + \,_2F_1\left[\frac12,\frac34;\frac32;y^2\right] \times y \right)\,,
\label{v+-function}\\[2mm]
P&=&P^+ (y)\,=\,P_0^+\,\frac{\left(1+y\right)^{3/4}}{\left(1-y\right)^{1/4}}
\,\,v^+ (y)\,,
\label{looking}
\end{eqnarray}
where $v_0^+$ is a dimensionless integration constant and $P_0^+$ is an integration constant with dimensions of (length)$^2$. Although $P_0^+$ sets the scale of the entire internal metric \eqq{eight}, we will show below that it can be completely eliminated from the full, eleven-dimensional metric once the warp factor is included. Nevertheless, we will need to fix the precise value of $P_0^+$ in order to ensure the same  value of the dual gauge coupling for all solutions. 

Given $y_0$, the condition \eqq{condition} fixes $v_0^+$ and vice versa. Hence we will think of $y_0$ as the parameter labelling the different solutions in the $\mathbb{B}_{8}^+$ family. The presence of $1-y^2$ in the denominators of the expressions above indicates that regular solutions correspond to $y_0\in\left[-1,1\right)$ or equivalently to $v_0^+\in\left(-v_c,v_c\right]$, where 
\be
v_c=\frac{\Gamma\left[1/4\right]^2}{\sqrt{8\pi}}
\ee
is the value of $v_0^+$ that can be read off from \eqq{v+-function} by setting $v=2$ and taking the limit $y^2 \to 1$. Looking at \eqq{looking}  and \eqq{looking2}, and noting that $v\geq 2$,   we see that in order for $P$ and $e^g$ to be positive we must have $P_0^+>0$ and $k<0$. As we will see, the negative sign of $k$ is consistent with the fact that the 
$\mathbb{B}_{8}^+$ family contains a flow to the OP solution \eqq{OPsolution}. 

As we mentioned above, the UV corresponds to the region $y\to1$, in which the behavior of the metric is universal for the entire family. In this region we can integrate the  change of coordinates \eqq{ry} to leading order to obtain 
\be
Q_k\left(1-y\right)^{1/4}r=2^{7/4}P_0^+ \,.
\ee
With this result we can write the transverse metric \eqq{eight}  at leading order as
\begin{equation}
\dd s_8^2\,\propto\,\dd r^2+\frac12 \, r^2\Bigg[  \dd\Omega_4^2+\frac12\left[\left(E^1\right)^2+\left(E^2\right)^2\right]\Bigg]+
\left(\frac{4P_0^+}{Q_k\left(v_0^++v_c\right)}\right)^2\left(E_3\right)^2\,.
\end{equation}
Since the size of the $E_3$ circle becomes constant,  we recognize this as the uplift of the D2-brane metric whose internal space in ten dimensions, given above between square brackets, is precisely the squashed $\mathbb{CP}^3$ at the NK point, corresponding to $\alpha^2=1/2$ in \eqq{CP3metric}. Given our parametrization of the dilaton in (\ref{runningD2}), in order for the solution to asymptote to the D2-brane solution (\ref{D2brane}) with the correct normalization  of the gauge coupling we must impose the boundary condition $e^{\Lambda}\to1$.  This fixes the dimensionful constant $P_0^+$ to the value
\begin{equation}
\label{P0choice}
P_0^+\,=\,\frac{Q_k^2\,\left(v_0^++v_c\right)}{4}\,,
\end{equation}
which, in particular, depends on $y_0$ through $v_0^+$. Since $\psi$ has period $4\pi/k$, this is equivalent to normalizing the asymptotic radius of the M-theory circle in the eight-dimensional transverse metric to the usual result
\be
R_{(11)} = \frac{2Q_k}{k} = g_s \ell_s \,.
\ee 
Note that $e^{\Lambda}\to1$ actually implies that we are setting $g_s=1$. Nevertheless, we will keep explicit factors of $g_s$ in our formulas in order to facilitate  comparison with the literature. 

For the $\mathbb{B}_{8}^+$ family we have $v_0^+\ne v_c$ and so the IR is located at $y \to y_0$, where the geometry ends. In a suitable radial coordinate $\rho$ defined through 
\be
4P_0^+\left(y-y_0\right)=\left(1-y_0\right)^{5/4}\left(1+y_0\right)^{1/4}\rho^2 \,, 
\ee
the transverse  metric at small $\rho$ approaches 
\begin{equation}
\label{gappedIR}
\dd s_8^2\,=\,\dd\rho^2+\frac14\rho^2\left[\left(E^1\right)^2+\left(E^2\right)^2+\left(E^3\right)^2\right]+\frac{2P_0^+\left(1+y_0\right)}{\left(1-y_0^2\right)^{1/4}}\,\dd\Omega_4^2\,.
\end{equation}
Since the $E^i$ describe a three-sphere fibration over ${\rm S}^4$, we find that in the IR the metric approaches locally $\mathbb{R}^4\times{\rm S}^4$, where the four-sphere has a finite radius squared proportional to $P_0^+$. However, solving for the warp factor with this transverse space we find an IR singularity, since near $y_0$ we have that 
\begin{equation}
H\,=\,\frac{Q_c|Q_k|\left(1-y_0\right)^{7/4}}{4\left(P_0^+\right)^3\left(1+y_0\right)^{5/4}}\left[\frac{1}{y-y_0}+\frac{1}{8\left(1-y_0^2\right)}\,\log\left(y-y_0\right)+\mathcal{O}\left(y-y_0\right)^0\right]\,,
\end{equation}
which diverges as $y\to y_0$. A singularity in the warp factor is also present for the rest of the solutions that we will discuss in this section. We will see in subsequent sections that this singularity can be removed by turning on appropriate additional components of  $F_4$ corresponding to fractional M2-branes \cite{Cvetic4}. 

\subsection{$\mathbb{B}_{8}^\infty$ solution}
\label{binfsol}

In the particular case $v_0^+=v_c$, corresponding to $y_0=-1$, after changing  variables through 
\be
P_0^+\,\left(1+y\right)^{3/4}=\frac{9}{2^{11/4}\times5 }\,\rho^{2} \,, 
\ee
one discovers that the  transverse space in the IR corresponds to the OP solution
\begin{equation}
\dd s_8^2\,=\,\dd \rho^2+\frac{9}{20}\,\rho^2\left[\dd\Omega_4^2+\frac{1}{5}\,\left[\left(E^1\right)^2+\left(E^2\right)^2+\left(E^3\right)^2\right]\right]\,,
\end{equation}
since one can recognize the metric inside the square brackets as the squashed seven-sphere. The full  geometry was denoted $\mathbb{B}_{8}^\infty$ in \cite{Hashimoto} and its significance had been overlooked in studies prior to this reference. It interpolates between the theory on the D2-branes on the squashed $\mathbb{CP}^3$ and the OP fixed point, so it can be seen as an irrelevant deformation of the OP CFT whose UV completion is a SYM-CSM  theory.  

\subsection{$\mathbb{B}_8^{\textrm{\tiny OP}}$ solution}

Remarkably, the OP fixed point also admits  a relevant deformation that can be solved for analytically. In our variables, the metric functions and dilaton are
\begin{equation}
e^{2f}\,=\,\frac95\,r^2\left[1-\left(\frac{r_0}{r}\right)^{5/3}\right],\quad e^{2g}\,=\,\frac{9}{25}\,r^2\left[1-\left(\frac{r_0}{r}\right)^{5/3}\right]^2,\quad e^{\Lambda}\,=\,\frac{3\,r}{5|Q_k|}\left[1-\left(\frac{r_0}{r}\right)^{5/3}\right]\,,
\end{equation}
with the radial direction ending at $r=r_0$, which plays a role analogous to that of $P_0^+$ in the $\mathbb{B}_8^+$ family. Changing coordinates from $r$ to $\varrho$ through 
\be
20|Q_k|r=3\varrho^2 
\ee
we see that this solution corresponds to the original Spin(7) manifold of \cite{Bryant, Gibbons}, whose metric is
\begin{equation}\label{Spin7holonomy}
\dd s_8^2\,=\,\frac{\dd \varrho^2}{\left[1-\left(\frac{\varrho_0}{\varrho}\right)^{10/3}\right]}+\frac{9}{20}\,\varrho^2\,\dd\Omega_4^2+\frac{9}{100}\,\varrho^2\,\left[1-\left(\frac{\varrho_0}{\varrho}\right)^{10/3}\right]\left[\left(E^1\right)^2+\left(E^2\right)^2+\left(E^3\right)^2\right]\,.
\end{equation}
The UV of this flow is of course the OP fixed point while the IR, which lies at $\varrho=\varrho_0$, is precisely of the form (\ref{gappedIR}), with the four-sphere radius proportional to $\varrho_0$.

\subsection{$\mathbb{B}_8$ solution}

At the other end  of the allowed values for $v_0^+$, i.e.~when $v_0^+\to-v_c$, corresponding to $y_0 \to1$, the coordinate $y$ ceases to be appropriate to describe the geometry. Instead, in the original radial coordinate $r$  the solution takes the simple form
\begin{equation}
e^{2f}\,=\,\frac12\,\frac{r^2\left(r-2r_0\right)}{\left(r-r_0\right)}\,,\quad e^{2g}\,=\,\frac14\,\frac{r^2\left(r-2r_0\right)^2}{\left(r-r_0\right)^2}\,,\quad e^{\Lambda}\,=\,\frac{r_0}{2|Q_k|}\,\frac{r\left(r-2r_0\right)}{\left(r-r_0\right)^2}\,,
\end{equation}
where one has to assume again that $k<0$ and 
\be
\label{r0}
r_0 = 2|Q_k| \,,
\ee
so that $e^\Lambda \to 1$ asymptotically. Note that the space ends at $r=2\,r_0$. Again, $r_0$ plays a role analogous to that of $P_0^+$ in the $\mathbb{B}_8^+$. Uplifting to eleven dimensions we get the transverse space
\begin{equation}
\dd s_8^2\,=\,\frac{\left(r-r_0\right)^2\dd r^2}{r\left(r-2r_0\right)}+\frac{1}{2}r\left(r-r_0\right)\dd\Omega_4^2+\frac{1}{4}r\left(r-2r_0\right)\left[\left(E^1\right)^2+\left(E^2\right)^2\right]+\frac{r_0^2}{4}\frac{r\left(r-2r_0\right)}{\left(r-r_0\right)^2}\left(E^3\right)^2\,.
\end{equation}
This metric was dubbed $\mathbb{B}_8$ in \cite{Cvetic2, Cvetic3}. The geometry ends smoothly at $r=2r_0$ and has the same asymptotic behavior as the  
$\mathbb{B}_{8}^+$ family. 

\subsection{$\mathbb{B}_{8}^-$ family}

Pushing further the values of $y_0$ one arrives to the $\mathbb{B}_{8}^-$ family of metrics, as described in \cite{Cvetic2, Cvetic3}. In our radial coordinate, they are defined in the range 
\be
1<y\le y_0<\infty \,, 
\ee
where again $v(y_0)=2$. The functions in the solution can be written as
 \begin{eqnarray}
v&=&v^-\,=\,\frac{1}{\left(y^2-1\right)^{1/4}}\left(v_0^-+\frac{2}{\sqrt{y}} \,\,_2F_1\left[\frac14,\frac34;\frac54; \frac{1}{y^2} \right]\right)\,,\nonumber\\[2mm]
P&=&P^-\,=\,P_0^-\,\frac{\left(y+1\right)^{3/4}}{\left(y-1\right)^{1/4}}\,\,v^-\,,
\end{eqnarray}
where  $v_0^-\in\left(-\sqrt2\,v_c,\infty\right)$ is a dimensionless integration constant and $P_0^-$ is an integration constant with dimensions of (length)$^2$ that sets the scale of the entire internal metric \eqq{eight}. To fix the correct asymptotics, $e^{\Lambda}\to1$, we must choose
\begin{equation}\label{P0mchoice}
P_0^-\,=\,\frac{Q_k^2\,\left(v_0^-+\sqrt{2}\,v_c\right)}{4}\,.
\end{equation}
Both the UV and IR behavior of the $\mathbb{B}_{8}^-$ metrics, located respectively at $y\to1$ and $y\to y_0$, coincide with those of the $\mathbb{B}_{8}^+$ family.

\subsection{$\mathbb{B}_8^{\rm{conf}}$ solution}

In all the solutions above we assumed that $k<0$. We close this section with the case $k=0$. Changing coordinates through 
\be
\dd r=\left(1-\frac{\rho_0^4}{\rho^4}\right)^{-1/2}\dd \rho 
\ee
the  BPS solution takes the form 
\begin{equation}
e^{\Lambda}\,=\,1 \,, \qquad\qquad e^{2f}\,=\,\frac12\,\rho^2 \,, \qquad\qquad e^{2g}\,=\,\frac14\,\rho^2\,\left(1-\frac{\rho_0^4}{\rho^4}\right)\,.
\end{equation}
Since $k=0$ the RR two-form  vanishes. This translates into the fact that the M-theory circle is trivially fibered and hence the eight-dimensional transverse metric in eleven dimensions is a direct product of the form $\mbox{M}_7 \times \mbox{S}^1$, where $\mbox{M}_7$ is the ${\rm G}_2$-manifold found in \cite{Bryant, Gibbons}. The UV corresponds again to D2-branes on the NK $\cpthree$.  
It can be seen that in the IR, located  at $\rho\to\rho_0$, the local geometry approaches  $\mathbb{R}^3\times{\rm S}^4$, with a finite radius for the S$^4$. Again there is a singularity in the warp factor that can be cured with additional fluxes corresponding to fractional D2-branes \cite{Cvetic}, as we describe in the following sections.  

The uplift of this metric is very simple and its transverse  part reads
\begin{equation}\label{G2holonomy}
\dd s_8^2\,=\,\frac{\dd \rho^2}{\left(1-\frac{\rho_0^4}{\rho^4}\right)}+\frac{1}{2}\,\rho^2\,\dd\Omega_4^2+\frac{1}{4}\,\rho^2\,\left(1-\frac{\rho_0^4}{\rho^4}\right)\left[\left(E^1\right)^2+\left(E^2\right)^2\right]+ \ell_p^2 \, \dd \psi^2\,.
\end{equation}
The IR limit corresponds now to $\mathbb{R}^3\times{\rm S}^1\times{\rm S}^4$, since the circle that before was fibered over ${\rm S^2}$ to form the three-sphere in $\mathbb{R}^4$ is now trivial and remains of finite size in the IR. As we will see, this change in topology has dramatic consequences in the dual field theory.

\section{Adding Fractional Branes}
\label{adding}

The transverse geometries presented in the previous section are suitable to support D2-brane solutions in ten dimensions  or M2-brane solutions in eleven dimensions preserving $\mathcal{N}=1$ supersymmetry in three dimensions. However, the corresponding warp factors diverge in the IR, thus rendering the full metrics singular.  Fortunately, these singularities can be removed by the standard procedure of adding new fluxes to the system. This mechanism was dubbed ``transgression"  in \cite{Cvetic}.  As usual in this type of constructions, the new fluxes can be interpreted as resulting from the addition of fractional branes and can be chosen so that supersymmetry is preserved.  

We start by reviewing the transgression mechanism as used for instance in \cite{Cvetic}. Imagine that one starts with the solution for a D2-brane preserving $\mathcal{N}=1$ supersymmetry, that is, a solution of the form (\ref{D2brane}) where the transverse  space is a (non-compact) G$_2$-holonomy manifold.
Since manifolds with special holonomy are Ricci-flat the only equation that needs to be solved is that for the warp factor,
\begin{equation}
\Box h=0\,,
\end{equation}
with $\Box$ the Laplacian of  the seven-dimensional transverse metric. Now suppose that we modify the ansatz for the four-form to include a new piece
\begin{equation}
F_4\,=\,\dd^3x\wedge\dd \left(h^{-1}\right)+G_4\,,
\end{equation}
where $G_4$ is a closed four-form on the transverse seven-dimensional space. The equation of motion for the NSNS three-form is then solved provided we also turn on  
\begin{equation}
H_3\,=\,*_7\,G_4\,,
\end{equation}
where  the Hodge dual is taken with respect to the transverse metric. Closure of  $H_3$ then implies that  $G_4$ is harmonic with respect to the G$_2$-holonomy metric. With this ansatz, all the equations of motion and Bianchi identities are satisfied as long as the warp factor obeys the inhomogeneous  equation
\begin{equation}\label{warpmodified}
\Box h\,=\,-\frac{1}{24}\,G_4^2\,.
\end{equation}
The key point of this construction is that the transverse geometry is not modified.

The solutions that we are interested in include a non-zero RR two-form, since they correspond to deformation of the field theory by the addition of CS terms. This means that the transgression mechanism above must be generalized as follows. 

Suppose that we have a solution of type IIA supergravity preserving at least $\mathcal{N}=1$ supersymmetry with metric and dilaton given by \eqq{runningD2}, with a four-form given by \eqq{F4flux}, and with a non-zero $F_2$ with components only along the compact directions. We also assume that the dilaton depends only  on the non-compact coordinates in $\dd s_7^2$. These conditions are satisfied by the solutions that we discussed in Sec.~\ref{sing} and by all the solutions of  \cite{Loewy, Aharony, Ooguri, Conde, us}. Under these assumptions the only non-trivial equations to solve are those that determine the dilaton and the warp factor or, equivalently, 
$\Lambda$ and $h$, and the equation of motion for $h$ can be derived from that for $F_4$.  

Now we would like  to turn on additional fluxes with the purpose of resolving potential singularities as those that we encountered in the warp factor in Sec.~\ref{sing} or, more generally, in order to add fractional branes to the system. Consider therefore  the following modification of the fluxes
\begin{eqnarray}
H_3&=&\dd B_2\,,\nonumber\\[2mm]
F_4&=&\dd^3x\wedge\dd\left(h^{-1}\,e^{-\Lambda}\right)+\left(G_4+B_2\wedge F_2\right)\,,
\label{f4f4}
\end{eqnarray}
where $G_4$ is closed and the following duality condition on the transverse space is satisfied
\begin{equation}\label{Duality}
e^{\Lambda}\left(G_4+B_2\wedge F_2\right)\,=\,*_7\,H_3\,.
\end{equation}
Under these circumstances the metric $\dd s_7^2$ and the function $\Lambda$ are left unchanged by the addition of the new fluxes. The only modified equation is that for the warp factor $h$. This can be derived from the equation  for $F_4$ and  becomes inhomogeneous because it is sourced by the new fluxes. We emphasize that the only assumptions about  $F_2$ are that it does not contain components along non-compact directions and that it verifies its Bianchi identity. We will now implement this generalized transgression for the solutions of Sec.~\ref{sing}.

The first task is to construct closed forms $G_4$ and $H_3$ on the metric (\ref{SquashedCP3}) satisfying the duality condition (\ref{Duality}). We start from the most general left-invariant ansatz using the forms defined on the coset:
\begin{eqnarray}
\label{all}
B_2&=&b_X\,X_2+b_J\,J_2\,,\nonumber\\[2mm]
H_3&=&\dd B_2\,,\nonumber\\[2mm]
G_4&=&\dd \left(a_X\,X_3+a_J\,J_3\right)+q_c\,\left(J_2\wedge J_2-X_2\wedge J_2\right) \,. 
\end{eqnarray}
Note that $H_3$ must be exact because it must be closed and  $\cpthree$ has no non-trivial three-cycles ---see \eqq{betti}. In contrast, $G_4$, which is also closed, can contain a non-exact piece along the non-trivial four-cycle of $\cpthree$. This non-trivial flux is a constant with dimensions of (length)$^3$  that we have called $q_c$ and that, as we will see, is related to the number of fractional branes. The rest of the coefficients $b_X, b_J, a_X, a_J$ are  functions of the radial coordinate that we will determine below.  

The duality condition \eqq{Duality} leads to the following set of equations
\begin{eqnarray}
\label{Fluxes}
a_X'&=&0\,,\nonumber\\[2mm]
a_J'&=&e^{-\Lambda}\left(b_X+b_J\right)\,,\nonumber\\[2mm]
b_X'&=&2\,e^{-4f+2g+\Lambda}
\big( q_c+2\,a_J-Q_k\,b_J\big)\,,\nonumber\\[2mm]
b_J'&=&e^{-2g+\Lambda}\Big[ Q_k\left(b_J-b_X\right) +2\,a_J-q_c\Big]\,.
\end{eqnarray}
Given the first equation and the closure of $X_3$, we see that the term  $a_X X_3$ does not contribute to $G_4$, and therefore we will henceforth set $a_X=0$. In addition, the requirement that $G_4$ be normalizable in the UV implies the relation 
\begin{equation}
\label{trunc}
a_J\,=\,\frac{e^{2g}\big( Q_kb_J-q_c\big)-2e^{2f+g-\Lambda}\big(b_J+b_X\big)+e^{2f}\Big[q_c+Q_k\big(b_X-b_J\big)\Big]}
{2\big(e^{2f}+e^{2g}\big)}\,.
\end{equation}
Moreover, the equation that is obtained by differentiating \eqq{trunc} is automatically satisfied by virtue of the equations for $b_J, b_X$ in \eqq{Fluxes}. We thus conclude that the system \eqq{Fluxes} can be reduced to two equations for the two functions $b_J$ and $b_X$.

Finally, as anticipated, the equation for the warp factor acquires  additional terms due to the flux sources and reads 
\begin{equation}\label{BPSwarp}
H'=\left(e^\Lambda\,h\right)' = -e^{2\Lambda-4f-2g}
\Big[Q_c+Q_k\,b_J \left(b_J- 2 b_X\right)+ 2 q_c \left(b_X- b_J\right) + 4 a_J\left(b_X+ b_J\right)\Big]\,,
\end{equation}
where we recall that the eleven- and the ten-dimensional warp factors are related through \eqq{HH}. 
Remember that the equations (\ref{BPSsystem}) for the background  are not modified by the new sources. This will allow us in the next section to solve the system sequentially: First we will solve for the background functions (\ref{BPSsystem}), then we will use that solution in (\ref{Fluxes}) and we will solve for the fluxes, and finally we will integrate the warp factor (\ref{BPSwarp}). As we will see, in some cases we will be able to find fully explicit analytic solutions.

Again there is a correspondence between our functions and the ones used in \cite{Cvetic2} (see also \cite{Hashimoto}) to construct a self-dual four-form in the eight-dimensional Spin(7) space. Specifically, the functions $u_i$ used in \cite{Cvetic2} are given by\footnote{As explained below \eqq{abc} we must set $Q_k=-1$ in order to reproduce \cite{Cvetic2} exactly.}
\begin{eqnarray}
u_1&=&4\,e^{2\Lambda-4f}\,\big(2a_J-Q_k\,b_J+Q_k^2\,q_c\big)\,,\nonumber\\[4mm]
u_2&=&2\,Q_k\,e^{2\Lambda-2f-2g}\Big[Q_k\left(b_X-b_J\right)-2a_J+q_c\big]\,,\nonumber\\[4mm]
u_3&=&-\frac{2e^{\Lambda-2f-g}}{Q_k}\big(b_X+b_J\big)\,.
\end{eqnarray}

In order to interpret the additional fluxes as fractional branes we need to properly quantize them. From the different notions of charge that may be defined in supergravity \cite{Marolf:2000cb}, the one that is quantized and conserved and counts the number of branes is the Page charge. We begin with the D2-brane charge. Following \cite{Hashimoto} we compute the number of D2-branes, which sets the rank of the dual gauge group, as
\begin{equation}
N\,=\,\frac{1}{2\kappa_{10}^2T_{{\rm D}2}}\int_{\mathbb{CP}^3}\left(-*F_4-B_2\wedge F_4+\frac12B_2\wedge B_2\wedge F_2\right)\,.
\end{equation}
Note that in the presence of the additional fluxes this equation replaces (\ref{chargenofrac}), since it is not just $*F_4$ but the full integrand above that is a closed form. Nevertheless, the result is the same relation \eqq{Qc} between $Q_c$ and $N$,  thus confirming that the new fluxes have not modified the number of non-fractional D2-branes. 

In the case of the D6-brane charge measured by the flux of $F_2$ through the $\mathbb{CP}^1$, equation  (\ref{chargenofrac}) with $p=6$ is unmodified by the new fluxes, since $F_2$ is closed. Hence the relation \eqq{Qk} between $Q_k$ and $k$ is also unmodified. 

Finally, the new fluxes induce D4-brane charge that is interpreted as $M$ fractional D2-branes. The quantization condition reads
\begin{equation}
\bar M \,=\,\frac{1}{2\kappa_{10}^2T_{{\rm D}4}}\int_{\mathbb{CP}^2}\left(F_4-B_2\wedge F_2\right)\,=\,\frac{1}{2\kappa_{10}^2T_{{\rm D}4}}\int_{\mathbb{CP}^2}G_4\,,
\end{equation}
where 
\be
\bar M = M-\frac{k}{2} \,.
\ee
In the coordinates introduced in \eqq{introduced} the $\mathbb{CP}^2$ four-cycle is characterized by $\theta=\varphi=\pi/2$,  so we get the relation
\begin{equation}
q_c\,=\,\frac{3\pi\ell^3_sg_s}{4} \,.
\end{equation}
Here $M$ represents the shift in the gauge group due to the fractional branes, while the $k/2$ shift was argued in \cite{Hashimoto2} to be needed to account for the Freed--Witten anomaly. We thus expect  the gauge group of the dual quiver to be ${\rm U}(N)_k\times{\rm U}(N+M)_{-k}$, where the subindices indicate the CS levels. In the next section we will construct the regular backgrounds that are dual to this theory.

\section{Regular Flows}
\label{regular}

We will now solve the equations that we introduced in the previous section in order to obtain regular geometries.  Recall that we must solve for two fluxes $b_J, b_X$ in \eqq{Fluxes} and for the warp factor $H$ in \eqq{BPSwarp}.
The dependence on the different charges can be factored out of the equations by writing them in terms of four dimensionless functions $\mathcal{B}_J, \mathcal{B}_X, \mathcal{A}_J$ and  $\mathcal{H}$ defined through
\begin{eqnarray}
\label{wesee}
b_J&=&-\frac{2q_c}{3|Q_k|}-\frac{\left(4q_c^2+3Q_c|Q_k|\right)^{1/2}}{3|Q_k|}\,\mathcal{B}_J\,,\nonumber\\[2mm]
b_X&=&\frac{2q_c}{3|Q_k|}+\frac{\left(4q_c^2+3Q_c|Q_k|\right)^{1/2}}{3|Q_k|}\,\mathcal{B}_X\,,\nonumber\\[2mm]
a_J&=&-\frac{q_c}{6}+\left(4q_c^2+3Q_c|Q_k|\right)^{1/2}\,\mathcal{A}_J\,,\\[2mm]
H&=&\frac{\left(4q_c^2+3Q_c|Q_k|\right)}{P_0^3}\,\mathcal{H}\,.\nonumber
\end{eqnarray}
Note that, although the constant terms in $b_J$ and $b_X$ combine to give a closed form that does not contribute to $H_3$, they do contribute to $B_2$. In the expression for the warp factor, by $P_0$ we mean $P_0^\pm$ for the $\mathbb{B}_{8}^\pm$ family and the corresponding analogous scale for the other metrics discussed in Sec.~\ref{sing}. At this point we can already see why these scales of the internal metric could be eliminated from the full, eleven-dimensional solution. Indeed, we see from \eqq{wesee} that $H \sim P_0^{-3}$ and from \eqq{eight} that $\dd s_8^2 \sim P_0$. As a consequence, $P_0$ cancels out in the $H^{1/3} \dd s_8^2$  term of the eleven-dimensional metric \eqq{M2metric}, and its contribution to the first term can be eliminated by rewriting the metric in  terms of rescaled gauge theory coordinates defined through 
\be
\label{resc}
\tilde x^\mu = \frac{P_0}{\left(4q_c^2+3Q_c|Q_k|\right)^{1/3}}\,\,   x^\mu \,. 
\ee
It follows that $P_0$ also cancels out in the ten-dimensional metric and dilaton, since these are directly read off from the eleven-dimensional metric. The RR-forms are also independent of $P_0$, since all the components \eqq{all} are manifestly $P_0$-independent, and the same rescaling \eqq{resc}  eliminates $P_0$ from the first term in $F_4$ in \eqq{f4f4}.

Substituting \eqq{wesee} in \eqq{Fluxes}  we find that  the dimensionless functions obey the equations 
\begin{eqnarray}
\label{obey}
\mathcal{B}_J'&=&\frac{6\mathcal{A}_J+\mathcal{B}_J+\mathcal{B}_X}{\left(v-2\right)\left(y-1\right)}\,,\nonumber\\[2mm]
\mathcal{B}_X'&=&\frac{2\left(v-2\right)\left(\mathcal{B}_J-6\mathcal{A}_J\right)}{v^2\left(y-1\right)\left(1+y\right)^2}\,,\\[2mm]
\mathcal{H}'&=&\frac{\mathcal{B}_J\left(\mathcal{B}_J+2\mathcal{B}_X\right)+12\mathcal{A}_J\left(\mathcal{B}_J-\mathcal{B}_X\right)-3}{36\left(1-y\right)^{1/4}\left(1+y\right)^{5/4}\left(v-2\right)^2}\nonumber\,,
\end{eqnarray}
where $\mathcal{A}_J$ is understood to be given by (\ref{trunc}) as
\begin{equation}
\mathcal{A}_J\,=\,\frac{\left(1+y\right)v^2-y\,v-2}{6\left(y+2\right)v-12}\,\mathcal{B}_J-\frac{v\left(1+y\right)\left(1+v\right)}{6\left(y+2\right)v-12}\,\mathcal{B}_X\,.
\end{equation}
$\mathcal{B}_J, \mathcal{B}_X, \mathcal{A}_J$ and $\mathcal{H}$ are functions only of $y$ and a given solution is labelled only by the parameter $y_0$, since all the dependence on the charges has been factored out. This makes these functions ideally suited for numerical integration. In order to do so, we first solve \eqq{obey} perturbatively around the IR and around the UV. 

\subsection{$\mathbb{B}_8^+$ family}

In the IR, defined by the condition $v(y_0)=2$, we find
\begin{eqnarray}
\mathcal{B}_J&=&1-\frac{1}{2\left(1-y_0^2\right)}\left(y-y_0\right)+\frac{2-3y_0}{8\left(1-y_0^2\right)^2}\left(y-y_0\right)^2+\mathcal{O}\left(y-y_0\right)^3\,, \nonumber\\[2mm]
\mathcal{B}_X&=&1-\frac{3}{4\left(1-y_0^2\right)^2}\left(y-y_0\right)^2+\mathcal{O}\left(y-y_0\right)^3 \,, \\[2mm]
\mathcal{H}&=&\mathcal{H}_{\text{\tiny{IR}}}-\frac{7}{48\left(1+y_0\right)^3\left(1-y_0^2\right)^{1/4}}\left(y-y_0\right)-\frac{77(y_0-2)}{576\left(1+y_0\right)^3\left(1-y_0^2\right)^{5/4}}\left(y-y_0\right)^2+\mathcal{O}\left(y-y_0\right)^3 \,,
\nonumber
\end{eqnarray}
where we have already imposed regularity of the warp factor, which fixes the integration constants in $\mathcal{B}_J$ and $\mathcal{B}_X$. The only undetermined constant in the IR expansion is $\mathcal{H}_{\text{\tiny{IR}}}$, which will be fixed in the full numerical solution by requiring  D2-brane asymptotics in the UV with the correct normalization. 

\begin{figure}[t]
\begin{center}
\begin{subfigure}{.45\textwidth}
\includegraphics[width=\textwidth]{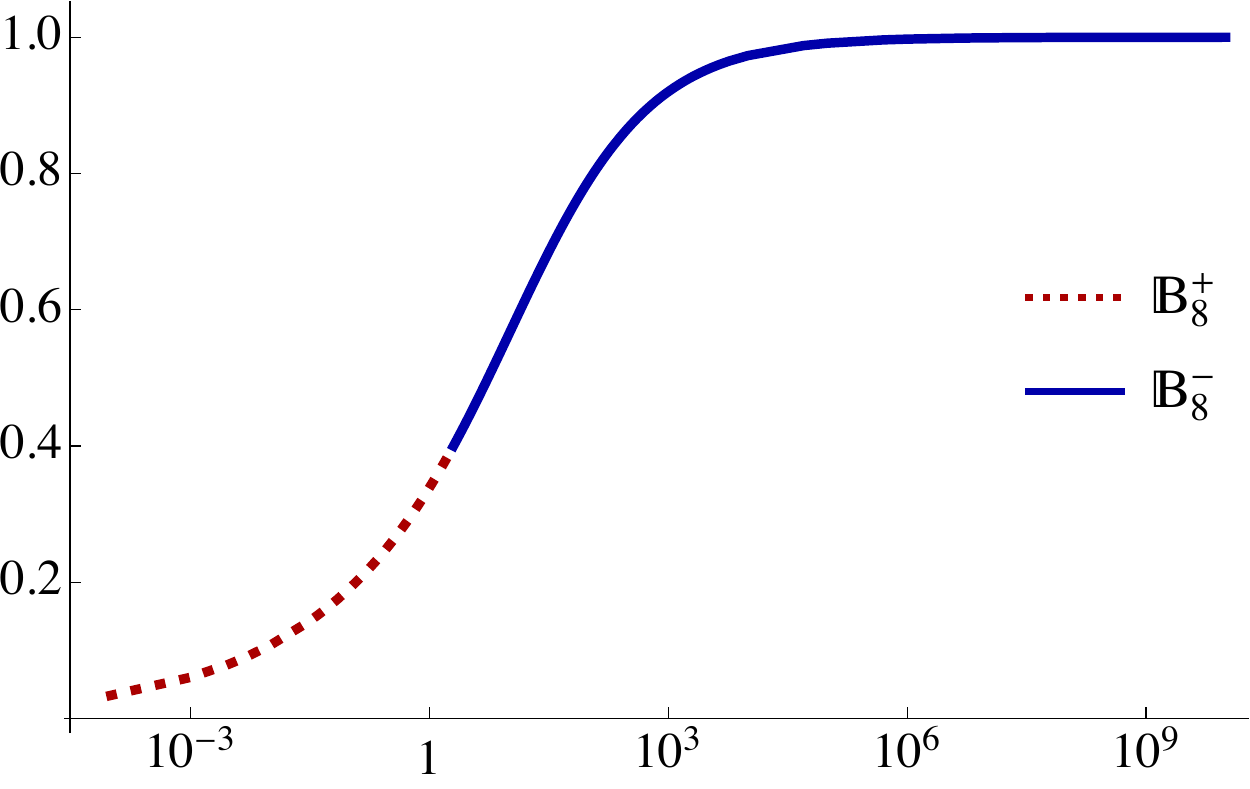} 
\put(-230,75){$b_0$}
\put(-100,-15){$y_0+1$}
\end{subfigure}\hfill
\begin{subfigure}{.45\textwidth}
\includegraphics[width=\textwidth]{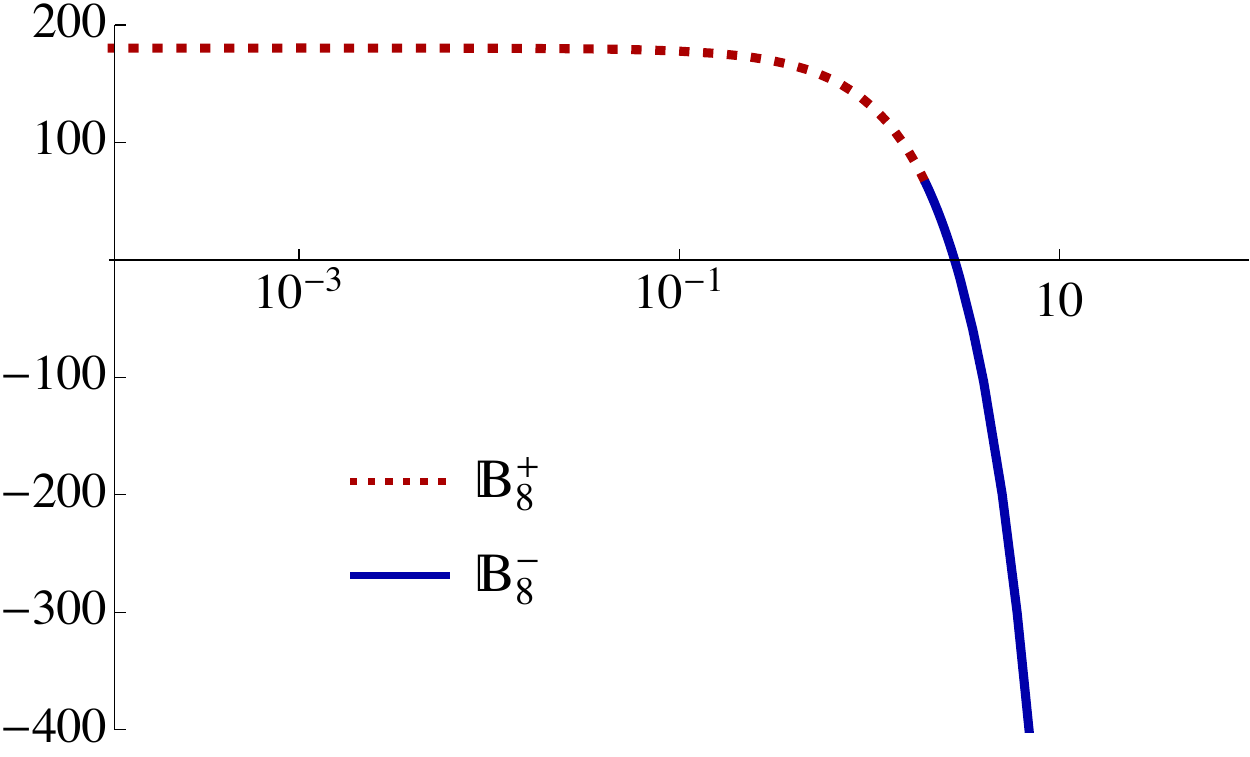} 
\put(-230,70){$b_4$}
\put(-100,-15){$y_0+1$}
\end{subfigure}

\caption{\small Values of the UV parameters $b_0$ (left) and $b_4$ (right) from the numerical integration.}\label{fig.numparam}
\end{center}
\end{figure}

\begin{figure}[t]
\begin{center}
\includegraphics[width=.55\textwidth]{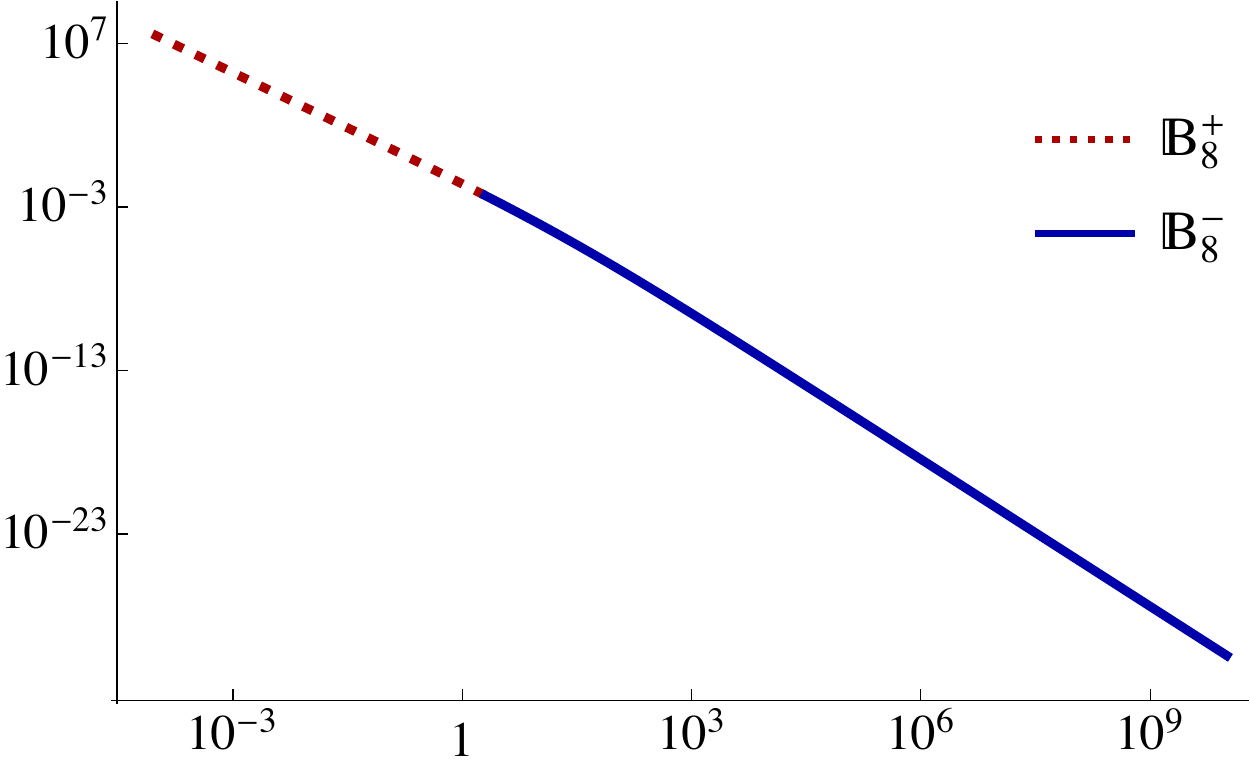} 
\put(-280,50){\rotatebox{90}{$64 \, \mathcal{H}_{\text{\tiny{IR}}} / (w_0^{\pm})^3 $}}
\put(-135,-15){$y_0+1$}
\caption{\small Values of the parameter $\mathcal{H}_{\text{\tiny{IR}}}$ from the numerical integration.}\label{fig.numHIR}
\end{center}
\end{figure}

In the UV, located at $y\to1$, we find  the expansions
\begin{eqnarray}\label{UVexpansions}
\mathcal{B}_J&=&b_0\left[1+\frac{2^{9/4}}{w_0^+}\left(1-y\right)^{1/4}+\frac{2^{7/2}}{\left(w_0^+\right)^2}\left(1-y\right)^{1/2}-\frac{2^{19/4}}{\left(w_0^+\right)^3}\left(1-y\right)^{3/4}+\frac{b_4}{\left(w_0^+\right)^4}\left(1-y\right)\right.\nonumber\\[2mm]
&&\left.\qquad+\mathcal{O}\left(1-y\right)^{5/4}\right]\,,
\nonumber\\[2mm]
\mathcal{B}_X&=&b_0\left[1+\frac{2^{9/4}}{w_0^+}\left(1-y\right)^{1/4}+\frac{3\times2^{5/2}}{\left(w_0^+\right)^2}\left(1-y\right)^{1/2}+\frac{2^{23/4}}{\left(w_0^+\right)^3}\left(1-y\right)^{3/4}-\frac{\left(128+\frac{b_4}{2}\right)}{\left(w_0^+\right)^4}\left(1-y\right)\right.\nonumber\\[2mm]
&&\left.\qquad+\mathcal{O}\left(1-y\right)^{5/4}\right] \,,\\[2mm]
\mathcal{H}&=&\mathcal{H}_{\text{\tiny{UV}}}+\frac{\left(1-b_0^2\right)}{15\times2^{3/4}\left(w_0^+\right)^2}\left(1-y\right)^{5/4}+\frac{2^{3/2}\left(1-2b_0^2\right)}{9\left(w_0^+\right)^3}\left(1-y\right)^{3/2}+\mathcal{O}\left(1-y\right)^{7/4}\,,\nonumber
\end{eqnarray}
with $w_0^+=\left(v_0^++v_c\right)\in\left(2v_c,0\right)$. The undetermined constants in the UV are thus $b_0, b_4$ and 
$\mathcal{H}_{\text{\tiny{UV}}}$. The latter must vanish in order to have the correct D2-brane asymptotics in the decoupling limit, i.e.~in order for $H\to0$ in the UV. Through the numerical integration, this requirement fixes the value of $\mathcal{H}_{\text{\tiny{IR}}}$. Once this is done there is a unique solution for each value of $y_0$ and the UV constants $b_0, b_4$ can be simply read off from the solution. The result is displayed in Fig.~\ref{fig.numparam}, whereas  Fig.~\ref{fig.numHIR} shows the IR value of the warp factor. The full solution is perfectly regular despite the fact that $\mathcal{H}_{\text{\tiny{IR}}}$ diverges as $y_0\to-1$. 

We see from Fig.~\ref{fig.numparam}(left) that there is a one-to-one correspondence between $y_0$ and the values of $b_0$ in the interval $(0,1)$. This is a nice consistency check of the fact that $y_0$ is related to the difference between the gauge couplings of the two gauge groups. The reason is that varying $b_0$ corresponds to varying the UV asymptotic flux of the NSNS two-form through the $\mathbb{CP}^1\subset\mathbb{CP}^3$. Since this asymptotic flux is expected to specify the difference between the gauge theory couplings \cite{Hashimoto}, the fact that $b_0$ can be mapped to $y_0$ in a one-to-one manner supports the idea that the family of theories under consideration are indeed parametrized by the difference between the gauge couplings. 

Presumably, $b_4$ is related to the vacuum expectation value of some operator in the gauge theory.

\subsection{$\mathbb{B}_{8}^\infty$ solution}

When $y_0\to-1$ the IR expansions above are not well defined, reflecting the dramatic change in the IR, which in this case is a fixed point instead of a gapped phase. Indeed, we have that the fluxes are constant
\begin{equation}
b_J\,=\,-\frac{2q_c}{3|Q_k|}\,,\qquad\qquad\qquad b_X\,=\,\frac{2q_c}{3|Q_k|}\,,\qquad\qquad\qquad a_J\,=\,-\frac{q_c}{6}\,.
\end{equation}
Using this, it is easy to find the expansions for the warp factor. In the IR, around $y=-1$, we get
\begin{equation}
\mathcal{H}\,=\,\mathcal{H}_{\text{\tiny{IR}}}+\left(y+1\right)^{-9/4}\frac{5}{9\times2^{5/4}}\left[\frac53-\frac{13}{8}\left(y+1\right)+\frac{815}{1664}\left(y+1\right)^2+\mathcal{O}\left(y+1\right)^3\right]\,.
\end{equation}
Notice that the constant term $\mathcal{H}_{\text{\tiny{IR}}}$ is not the leading term in this case  and this causes the metric to be AdS. On the other hand, the UV expansion gives again D2-brane asymptotics, as can be obtained from the general expansion of the $\mathbb{B}_8^+$ family, specifying to $w_0^+=2v_c$ and $b_0=0$.

The only parameter to be found from the numerics is $\mathcal{H}_{\text{\tiny{IR}}}$ such that the warp factor has no constant piece in the UV. From our results we find $\mathcal{H}_{\text{\tiny{IR}}}\simeq-0.0087$.

\subsection{$\mathbb{B}_8^{\textrm{\tiny OP}}$ solution}

The RG flow that connects the OP fixed point to the gapped phase can also be solved for analytically. In terms of a dimensionless coordinate 
\be
\rho=\frac{r}{r_0} 
\ee
 the fluxes are simply
\begin{equation}
\mathcal{B}_J\,=\,\frac{1}{\rho^{1/3}}\,,\qquad\qquad\qquad \mathcal{B}_X\,=\,\frac{6\rho^{5/3}-1}{5\rho^2}\,,
\end{equation}
and the regular warp factor is
\begin{eqnarray}
\mathcal{H}&=&\frac{5}{243}\left[\frac{1}{\rho^2}-\frac{9}{\rho^{1/3}}-3\,\frac{\rho^{4/3}-\rho^{1/3}}{\rho^{5/3}-1}\right]\\[4mm]
&+&\frac{4\sqrt{2}}{81}\left(\sqrt{5+\sqrt{5}}\arctan\left[\frac{\sqrt{10+2\sqrt{5}}}{4\rho^{1/3}+1-\sqrt{5}}\right]+\sqrt{5-\sqrt{5}}\arctan\left[\frac{\sqrt{10-2\sqrt{5}}}{4\rho^{1/3}+1+\sqrt{5}}\right]\right) \,.\nonumber
\end{eqnarray}

\subsection{$\mathbb{B}_8$ solution}

In this case a complete analytic solution can be found. In terms of a dimensionless coordinate 
\be
\rho=\frac{r}{r_0} \,,
\ee
with $r_0$ given by \eqq{r0}, the  fluxes take the form
\begin{equation}
\mathcal{B}_J\,=\,\frac{2\left(\rho^4+\rho^3-4\,\rho+4\right)}{5\rho^3\left(\rho-1\right)}\,,\qquad\qquad\qquad\mathcal{B}_X\,=\,\frac{2\left(\rho^5-10\,\rho+8\right)}{5\rho^3\left(\rho-1\right)^2}\,,
\end{equation}
where one integration constant was fixed to have D2-brane asymptotics in the UV while the other two were fixed by regularity.  The M2-brane warp factor can be found again in closed form and is simply
\begin{equation}
\mathcal{H}\,=\,\frac{\left(1323\rho^6+924\rho^5+963\rho^4+510r^3-1340\rho^2-4340\rho+2800\right)}{47250\,\rho^9\left(\rho-1\right)^2}\,,
\end{equation}
which is perfectly regular at $\rho=2$. Notice that the boundary conditions have fixed all the integration constants, the only parameters being the quantized charges.

\subsection{$\mathbb{B}_8^-$ family}

For $y_0>1$ the equations admit expansions similar to those of the $\mathbb{B}_8^+$ family. Around the end of the geometry, imposing regularity, we find 
\begin{eqnarray}
\mathcal{B}_J&=&1+\frac{1}{2\left(y_0^2-1\right)}\left(y-y_0\right)+\frac{2-3y_0}{8\left(y_0^2-1\right)^2}\left(y-y_0\right)^2+\mathcal{O}\left(y-y_0\right)^3\,, \nonumber\\[2mm]
\mathcal{B}_X&=&1-\frac{3}{4\left(y_0^2-1\right)^2}\left(y-y_0\right)^2+\mathcal{O}\left(y-y_0\right)^3 \,, \\[2mm]
\mathcal{H}&=&\mathcal{H}_{\text{\tiny{IR}}}+\frac{7}{48\left(y_0+1\right)^3\left(y_0^2-1\right)^{1/4}}\left(y-y_0\right)-\frac{77(y_0-2)}{576\left(y_0+1\right)^3\left(y_0^2-1\right)^{5/4}}\left(y-y_0\right)^2+\mathcal{O}\left(y-y_0\right)^3 \,.
\nonumber
\end{eqnarray}
Again, we have $\mathcal{H}_{\text{\tiny{IR}}}$ as the only undetermined constant in the IR, which will be fixed in the numerics by the UV conditions. Similarly, for $y\to1$ we have expansions identical to those in (\ref{UVexpansions}) with the replacements $\left(1-y\right)\to\left(y-1\right)$ and $w_0^+\to w_0^-=\left(v_0^-+\sqrt{2}\,v_c\right)\in\left(0,\infty\right)$. The equations are solved using these expansions, with $\mathcal{H}_{\text{\tiny{UV}}}=0$ for D2-brane asymptotics, as boundary conditions. 

\subsection{$\mathbb{B}_8^{\rm{conf}}$ solution}

In this case it is convenient to change from the $\rho$ coordinate in (\ref{G2holonomy}) to a dimensionless coordinate 
\be
z=\frac{\rho}{\rho_0} \,. 
\ee
The fluxes regularizing the solution are 
\begin{eqnarray}
b_J&=&\frac{Q_c}{4q_c}+\frac{2q_c}{3\rho_0}\left[\frac{z\sqrt{z^4-1}-\left(3z^4-1\right)U(z)}{z^4-1}\right]\,,\nonumber\\[2mm]
b_X&=&-\frac{Q_c}{4q_c}-\frac{2q_c}{3\rho_0}\left[\frac{z\sqrt{z^4-1}-\left(3z^4-1\right)U(z)}{z^4}\right]\,,
\end{eqnarray}
where the dimensionless function $U$ is defined as 
\begin{equation}
U(z)\,=\,\int_1^z\left(\sigma^4-1\right)^{-1/2}\dd \sigma\,.
\end{equation}
The warp factor, both in ten and eleven dimensions, is given by  
\begin{equation}
\label{given}
h=H=\frac{128q_c^2}{9\rho_0^6}\int_z^\infty\left[\frac{2-3\sigma^4}{\sigma^3\left(\sigma^4-1\right)^2}+\frac{\left(4-9\sigma^4+9\sigma^8\right)U(\sigma)}{\sigma^4\left(\sigma^4-1\right)^{5/2}}+\frac{2\left(1-3\sigma^4\right)U(\sigma)^2}{\sigma^5\left(\sigma^4-1\right)^3}\right]\dd\sigma\,.
\end{equation}

\subsection{Range of validity}

We now turn to the determination of the range of validity of the supergravity solutions above. Since in the UV the dilaton goes to zero, the correct description is the ten-dimensional one. This one extends up to the UV scale at which the curvature ceases to be small in string units. The Ricci scalar of the ten-dimensional solutions grows in the UV as 
\begin{equation}
\ell_s^2 R\sim \ell_s^2\left(\frac{|Q_k|}{4q_c^2+3Q_c|Q_k|}\frac{r}{\left(1-b_0^2\right)}\right)^{1/2}\,.
\end{equation}
Requiring this to be small and translating to a gauge theory energy scale $U$ via the usual relation $U=r/\ell_s^2$ \cite{Itzhaki} we find the condition
\begin{equation}
U\ll\lambda\left(1+\frac{\bar{M}^2}{2N|k|}\right)\left(1-b_0^2\right)\,,
\end{equation}
where we recall that $\lambda$ is the 't Hooft coupling with dimensions of energy. We observe that the usual result $U\ll\lambda$ for the D2-branes gets  modified due to the presence of the fractional branes. We have included the dependence on $y_0$ through the coefficient $1-b_0^2$,  which vanishes as $y_0^{-1/2}$ when $y_0\to\infty$. This is a manifestation of the fact that, in the limit $y_0 \to \infty$, $Q_k$ must scale as $Q_k \sim k \sim y_0^{-1/2}$ in order to obtain a valid supergravity description. The origin of this scaling together with more details will be given in the next section.   

\begin{figure}[t]
\begin{center}
\includegraphics[width=.55\textwidth]{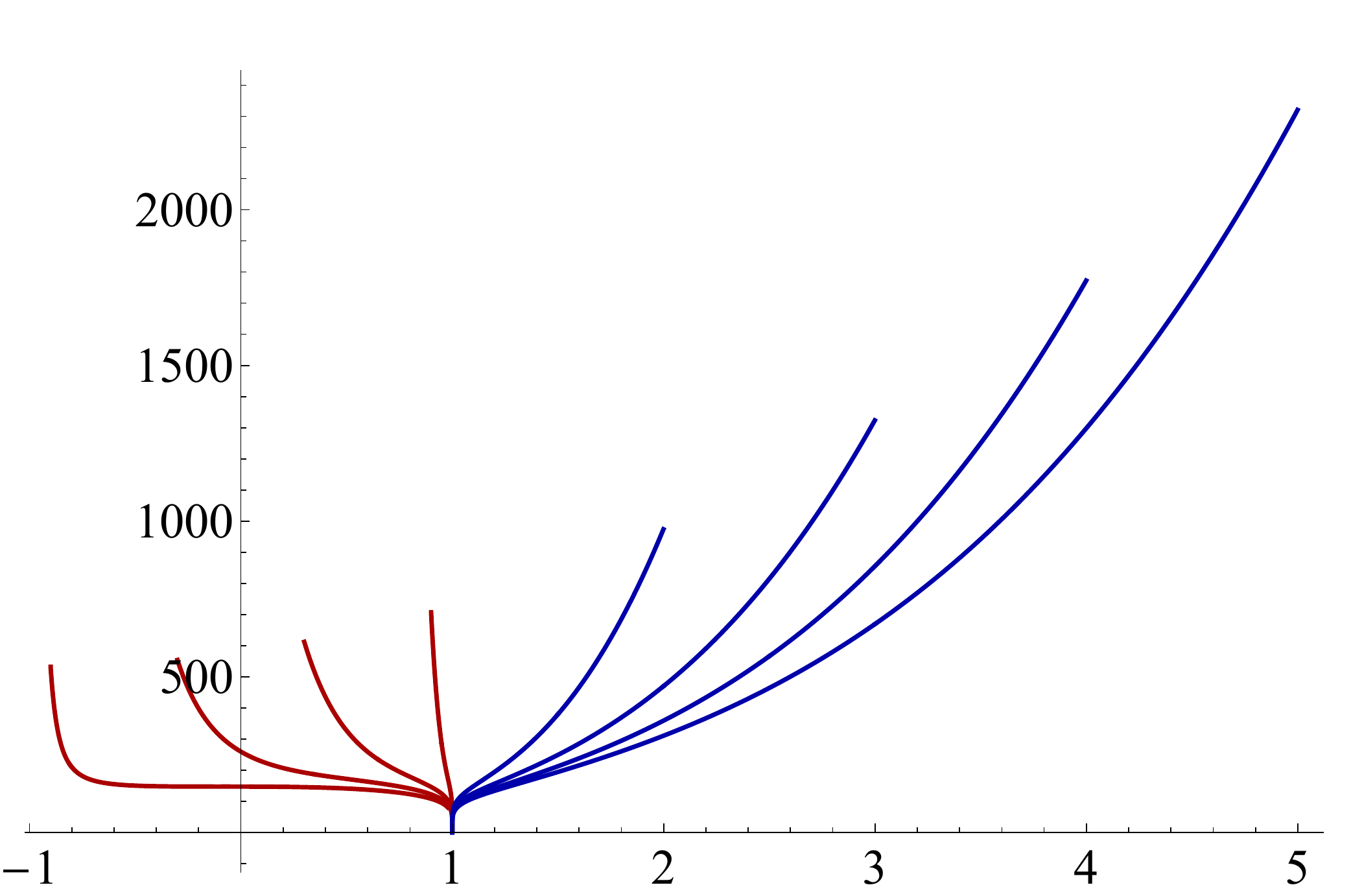} 
\put(-270,170){\rotatebox{0}{$\left(4q_c^2+3Q_c|Q_k|\right)^{2/3} K$}}
\put(-135,-15){$y$}
\caption{\small Kretschmann scalar 
$K=R_{\mu \nu \rho \sigma}  R^{\mu \nu \rho \sigma}$ as a function of $y$ for $\Bplus$ solutions (left, red curves) with $y_0=-0.9, -0.3, 0.3, 0.9$ from left to right, and for  $\Bminus$ solutions (right, blue curves) with $y_0= 2,3,4,5$ from left to right. We see that at $y=1$ all curves approach the same value since they all share the same UV asymptotics, whereas the curvature at the IR endpoint ($y=y_0$) diverges as $y_0^2$ as $y_0\to\infty$. }
\label{Kret}
\end{center}
\end{figure}

In the IR the ten-dimensional metrics are singular, so the correct description is given in terms of the eleven-dimensional solutions, in which the IR value of the Ricci scalar in units of the eleven-dimensional Planck length $\ell_p = g_s^{2/3} \ell_s$ is finite and scales  as
\begin{equation}
\ell_p^2\, R\sim\left(\frac{\bar{M}^2}{2}+N|k|\right)^{-1/3}\,.
\end{equation}
In order for this to be small we simply need to require that the combination 
\be
\frac{\bar{M}^2}{2}+N|k| \gg 1 \,.
\ee
For large $y_0$, however, the IR value of the Kretschmann scalar, $K=R_{\mu \nu \rho \sigma}  R^{\mu \nu \rho \sigma}$, shown in \Fig{Kret}, grows as 
\be
\ell_p^4 \, K \sim \left(\frac{\bar{M}^2}{2}+N|k|\right)^{-2/3} y_0^2  \,.
\ee
Thus in the limit $y_0 \to \infty$ we must impose the additional condition that 
\begin{equation}
|k|^6\left(\frac{\bar{M}^2}{2}+N|k|\right)\gg1\,,
\end{equation}
 where again we have assumed that $k \sim y_0^{-1/2}$.

\section{Limiting Dynamics}
\label{quasi}
In this section we will study the limits of the above metrics as $y_0\to \infty$ and as $y_0 \to -1$.  In the first  case the solution approaches $\Bconf$ everywhere except in the deep IR. In the second case the solution approaches the combination of the $\Binf$ flow followed by the $\Bop$ flow. In this sense the solutions with generic $y_0$ continuously interpolate between quasi-confining and quasi-conformal  dynamics. We will verify this with an explicit calculation of the quark-antiquark potential in Sec.~\ref{potential}. 

\subsection{Quasi-confining dynamics}
\label{quasiconfining}

Consider the limit $y_0\to \infty$ of the $\Bminus$ solutions. Expanding the functions of the internal metric for large $y_0$ we find 
\begin{eqnarray}\label{largey0}
e^{2f}&=&\frac{4\left(P_0^-\right)^2}{|Q_k|^2}\left(\frac{y+1}{y-1}\right)^{1/2}\left[1-\frac{\left(y^2-1\right)^{1/4}}{\sqrt{y_0}}+\mathcal{O}\left(y_0^{-1}\right)\right]\,,\nonumber\\[2mm]
e^{2g}&=&\frac{4\left(P_0^-\right)^2}{|Q_k|^2}\frac{1}{\left(y^2-1\right)^{1/2}}\left[1-\frac{2\left(y^2-1\right)^{1/4}}{\sqrt{y_0}}+\mathcal{O}\left(y_0^{-1}\right)\right]\,.
\end{eqnarray}
Performing the change of variables
\begin{equation}
y\,=\,\frac{\rho^4+\rho_0^4}{\rho^4-\rho_0^4}
\end{equation}
we see  that,  to leading order, we recover the confining metric (\ref{G2holonomy}) with an internal scale given by
\begin{equation}
\rho_0^2 \,=\, \frac{8 \left(P_0^-\right)^2}{|Q_k|^2}  \,.
\end{equation}
Given that $P_0^-$ was fixed by the UV condition $e^\Lambda\to1$ as in (\ref{P0mchoice}),  to leading order in $y_0$ we have 
\begin{equation}
\label{rho0}
\rho_0^2\,=\,2\,|Q_k|^2\,y_0\,.
\end{equation}
Note that, since $y_0 \to \infty$, $\rho_0$ seems to grow without bound. We may think of the limit in two (equivalent) ways. One is simply to keep all charges fixed as we take $y_0 \to \infty$ but to rescale the gauge theory coordinates as in \eqq{resc} with $P_0$ replaced by $\rho_0$, since this cancels all the dependence of the solution on $\rho_0$. The other is to keep the gauge theory coordinates fixed  but to scale $Q_k \sim y_0^{-1/2}$ as we take $y_0 \to \infty$. This is intuitive since we know that the 
 $\Bconf$ solution has $k=0$. By comparing with the analytic confining solution (\ref{G2holonomy}) it is possible to deduce how the parameters $b_0$, $b_4$ and $\mathcal{H}_{\text{\tiny{IR}}}$ must scale for large $y_0$, with the result
\begin{equation}
\left(1-b_0^2\right)\,\sim\,\frac{6K\left(-1\right)}{\sqrt{2}}\,y_0^{-1/2}\,,\qquad\qquad b_4\,\sim\,-2^{7/2}\,K\left(-1\right)\,y_0^{3/2}\,,\qquad\qquad \mathcal{H}_{\text{\tiny{IR}}}\,\sim\,\frac{h_{\rm conf}}{256}\,y_0^{-3/2}\,,
\end{equation}
where  $K(m)$ is the complete elliptic integral of the first kind and $h_{\rm conf}$ is  the IR value of the warp factor for the confining solution given by Eq.~(\ref{given}) with $z=1$. We have verified these scalings with our numerical solutions. One way to see that both ways of taking the limit are equivalent is to note that in both cases the $y_0$-dependent coefficient in front of the $\dd x_{1,2}^2$ term in \eqq{runningD2} attains a finite limit as $y_0 \to \infty$. 

In terms of the $\rho$ coordinate, the first correction in (\ref{largey0}) (the second term inside the square brackets) takes the form
\begin{equation}
\frac{\rho_0}{\rho}\frac{1}{\sqrt{y_0}\left(1-\frac{\rho_0^4}{\rho^4}\right)^{1/2}}\,.
\end{equation}
We see that, no matter how large $y_0$ is, this first correction  competes with the  leading term (the 1 in (\ref{largey0})) sufficiently close to $\rho_0$. This was expected because we know that, sufficiently deep in the IR, the $\Bminus$ and the $\Bconf$ metric differ dramatically from one another: in $\Bminus$ the M-theory circle shrinks to zero size whereas in $\Bconf$ it does not. The intuitive picture is therefore that, by taking $y_0$ large enough, one can make the 
$\Bminus$ and the $\Bconf$ metrics arbitrarily close to one another on an energy range that extends form the UV down to an IR scale arbitrarily close to the mass gap. Throughout this range the ${\rm S}^1$ of the internal metric has a constant and identical size in both metrics. Sufficiently close to the mass gap, however, the $\Bminus$ metric abruptly deviates from the $\Bconf$ metric and the internal ${\rm S}^1$ closes off. Presumably  this fast change of the size of the circle is related  to the fact that the curvature in the deep IR diverges as $y_0 \to \infty$, as shown in \Fig{Kret}. 

\subsection{Quasi-conformal dynamics}
\label{quasiconformal}

The $\Binf$ and the $\Bop$ solutions arise as two different limits of the $\Bplus$ metrics. If the limit $y_0=-1$ of the $\mathbb{B}_8^+$ is taken at fixed $y$ then the result is the $\mathbb{B}_8^\infty$ solution, as we saw in \Sec{binfsol}. 

Instead, if we first focus on the IR of $\mathbb{B}_8^+$ by expanding around $y-y_0$, so that we see the $\mathbb{R}^4\times{\rm S}^4$ region, and afterwards take the $y_0\to-1$ limit, then the $\mathbb{B}_8^{\textrm{\tiny OP}}$ metric is reproduced. 
Indeed, for the size of the four-sphere in the eight-dimensional transverse space we have in the strict IR 
\begin{equation}
e^{2f-\Lambda}\,=\,2^{3/4}\,P_0\,\left(y_0+1\right)^{3/4}+\mathcal{O}\left(y_0+1\right)^{7/4}\,.
\end{equation}
Comparing with the IR expansion for $\mathbb{B}_8^{\textrm{\tiny OP}}$ suggests the relation 
\begin{equation}\label{r0P0}
r_0\,=\,\frac{2^{3/4}P_0}{3|Q_k|}\left(y_0+1\right)^{3/4}\,.
\end{equation}
As in the previous subsection, we may take the limit in two ways, either by rescaling the gauge theory directions or by rescaling  $Q_k$. In the latter case, in order for $r_0$ to be finite, we must scale  $Q_k$ as 
$\left(y_0+1\right)^{-3/4}$ when $y_0\to-1$. Moreover, using this identification of parameters and integrating the change of coordinates (\ref{ry}) in the IR and around $y_0+1$ we get
\begin{eqnarray}
y-y_0&=&\left[\frac{4\left(r-r_0\right)}{3r_0}+\frac{2\left(r-r_0\right)^2}{9r_0^2}+\mathcal{O}\left(r-r_0\right)^3\right]\left(y_0+1\right)\nonumber\\[2mm]
&-&\left[\frac{5\left(r-r_0\right)}{6r_0}+\frac{5\left(r-r_0\right)^2}{6r_0^2}+\mathcal{O}\left(r-r_0\right)^3\right]\left(y_0+1\right)^2+\mathcal{O}\left(y_0+1\right)^3\,.
\end{eqnarray}
Finally, substituting this expansion together with (\ref{r0P0}) for the metric functions in the IR of the $\mathbb{B}_8^+$ family and taking the $y_0\to-1$ limit we arrive at
\begin{eqnarray}
e^{2f}&=&3r_0\left(r-r_0\right)+2\left(r-r_0\right)^2-\frac{\left(r-r_0\right)^3}{9r_0}+\mathcal{O}\left(r-r_0\right)^4\,,\nonumber\\[2mm]
e^{2g}&=&\left(r-r_0\right)^2-\frac{2\left(r-r_0\right)^3}{3r_0}+\mathcal{O}\left(r-r_0\right)^4\,,
\end{eqnarray}
which coincides, to this order, with the corresponding expansions for $\mathbb{B}_8^{\textrm{\tiny OP}}$.

\begin{figure}[t]
\begin{center}
\includegraphics[width=.58\textwidth]{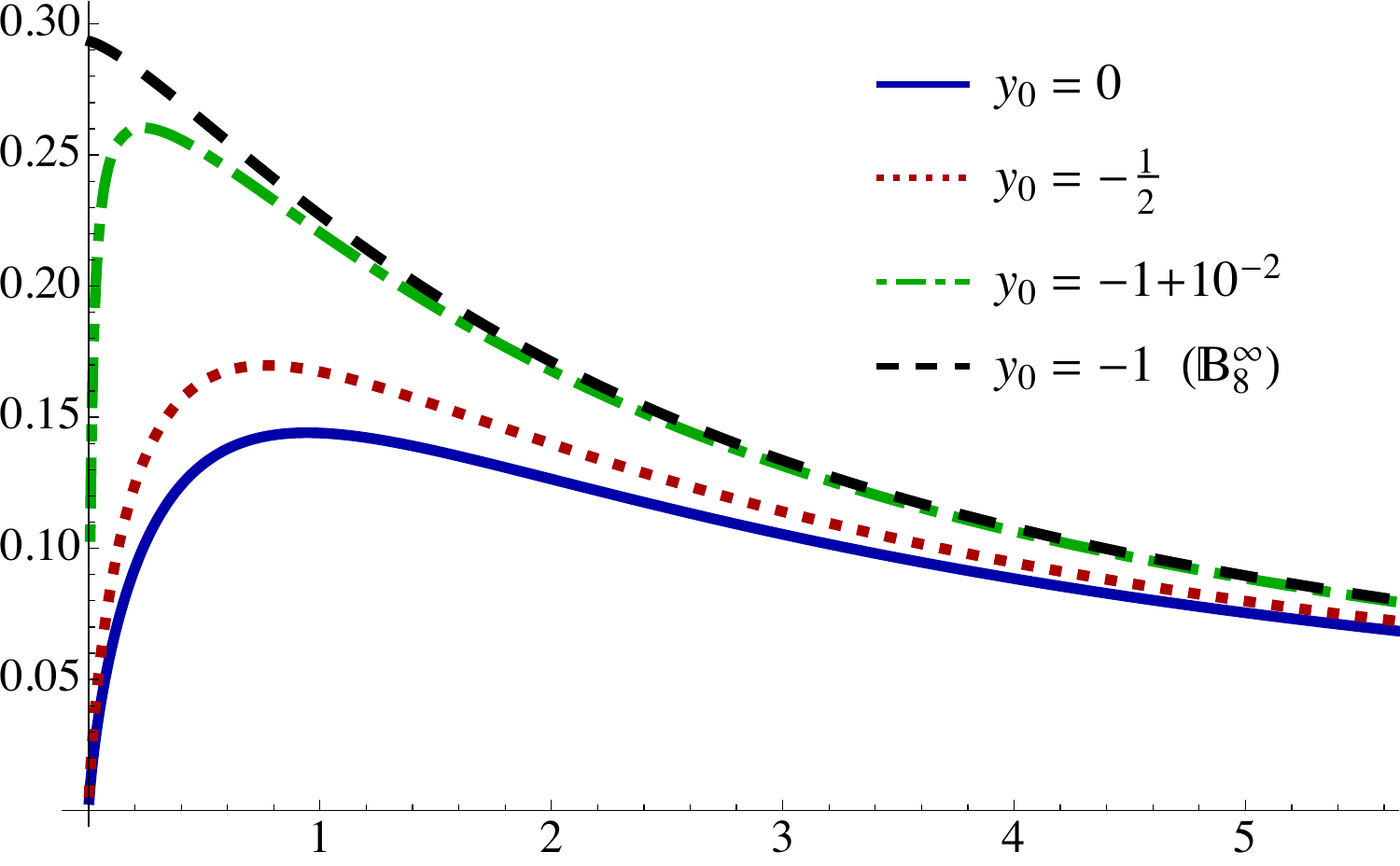} 
\put(-310,185){$\left( \frac{Q_k^6}{4q_c^2 + 3|Q_k| Q_c}\right)^{1/4} e^\Phi$}
\put(10,5){$M/M_0$}
\caption{\small Dilaton as a function of the energy scale for several $\Bplus$ solutions.}
\label{dil}
\end{center}
\end{figure}

The intuitive picture is therefore that $\Bplus$ solutions with $y_0 \gtrsim -1$  flow very close to the OP fixed point but eventually deviate from it and develop a mass gap. The mass scale $M_*$ at which the deviation occurs can be estimated from the behavior of the dilaton, which is plotted in \Fig{dil}. 
The mass scale $M$ on the horizontal axis is the mass of a membrane stretched from the bottom of the geometry until the position $y$ at which the dilaton  is evaluated (see \Sec{potential}). The normalization factor is 
\be
\label{MM}
M_0 = \frac{|Q_k|}{2\pi \ell_s^2}=\frac{\lambda \, k}{4\pi \, N} \,.
\ee
We define $M_*$ as the position of the maximum of each curve. 
We see that curves with $y_0 \to -1$ tend to the $\Binf$ curve but eventually deviate from it around the scale $M_*$ and approach zero at the end of the geometry, instead of approaching the OP value as $\Binf$ does.

\section{Quark-antiquark Potential}
\label{potential}

We will now present a computation of the potential between an external quark and an external antiquark separated by a distance $L$ in the gauge theory directions. In the string description this would be extracted from the action of a string hanging from the quark and the antiquark. Instead, in M-theory we must consider a hanging membrane. Although the membrane is asymptotically wrapped on the M-theory circle, namely on the ${\rm S}^1$ fiber of the internal geometry, as the membrane penetrates into the bulk geometry the circle wrapped by the membrane may vary. In particular, since the ${\rm S}^1$ fiber is  contractible inside the ${\rm S}^7$, the circle wrapped by the membrane may shrink to zero size at some value of the holographic coordinate, even if at that point the entire S$^7$ has finite size. All in all this means that, in order to find the membrane with the minimum energy, strictly speaking we would need to solve a problem involving partial differential equations for the membrane embedding  as a function of two worldspace intrinsic coordinates. Since this calculation is beyond the scope of this paper, we will perform a simpler one that nevertheless  is expected to capture the qualitative physics. We will therefore assume that the circle wrapped by the membrane is the ${\rm S}^1$ fiber at all values of the holographic coordinate. This reduces the problem to that of solving ordinary differential equations. 

An important point in the calculation is that, generically, the membrane action is UV divergent. We will renormalize away this divergence by subtracting the action of two disconnected membranes extending from the UV all the way to the IR end of the geometry. For all metrics except for $\Bconf$ this is in itself a physically acceptable configuration that competes with the connected configuration. In the case of $\Bconf$ the disconnected configuration is not a physically acceptable configuration to which the connected one can transition, but it can still be used as a mathematical well defined quantity that can be used  to regularize the membrane action.

\begin{figure}[t]
\begin{center}
\begin{subfigure}{.45\textwidth}
\includegraphics[width=\textwidth]{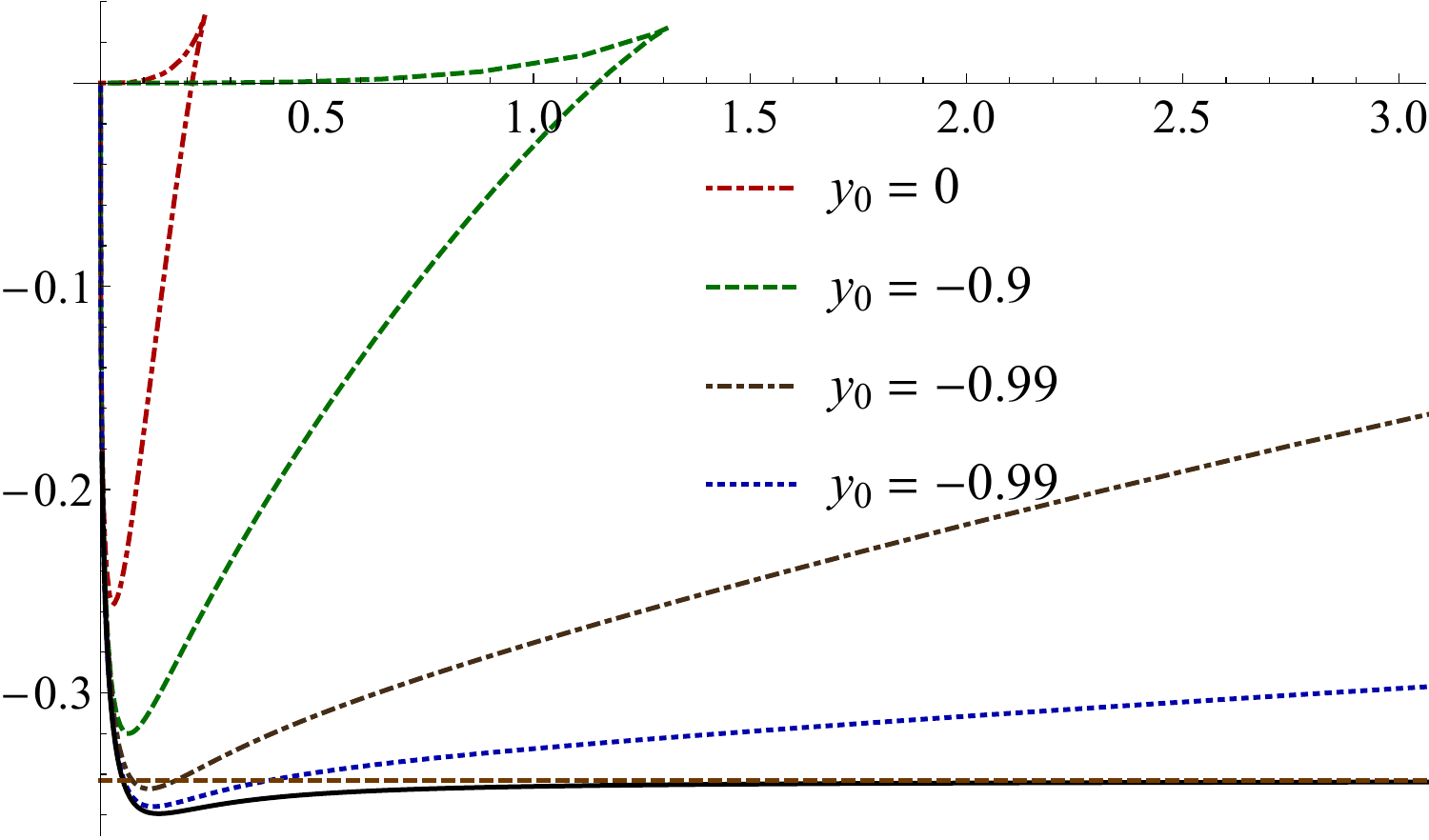} 
\put(-220,140){$V L / M_0 L_0$}
\put(-25,123){$L/L_0$}
\end{subfigure}\hfill
\begin{subfigure}{.45\textwidth}
\includegraphics[width=\textwidth]{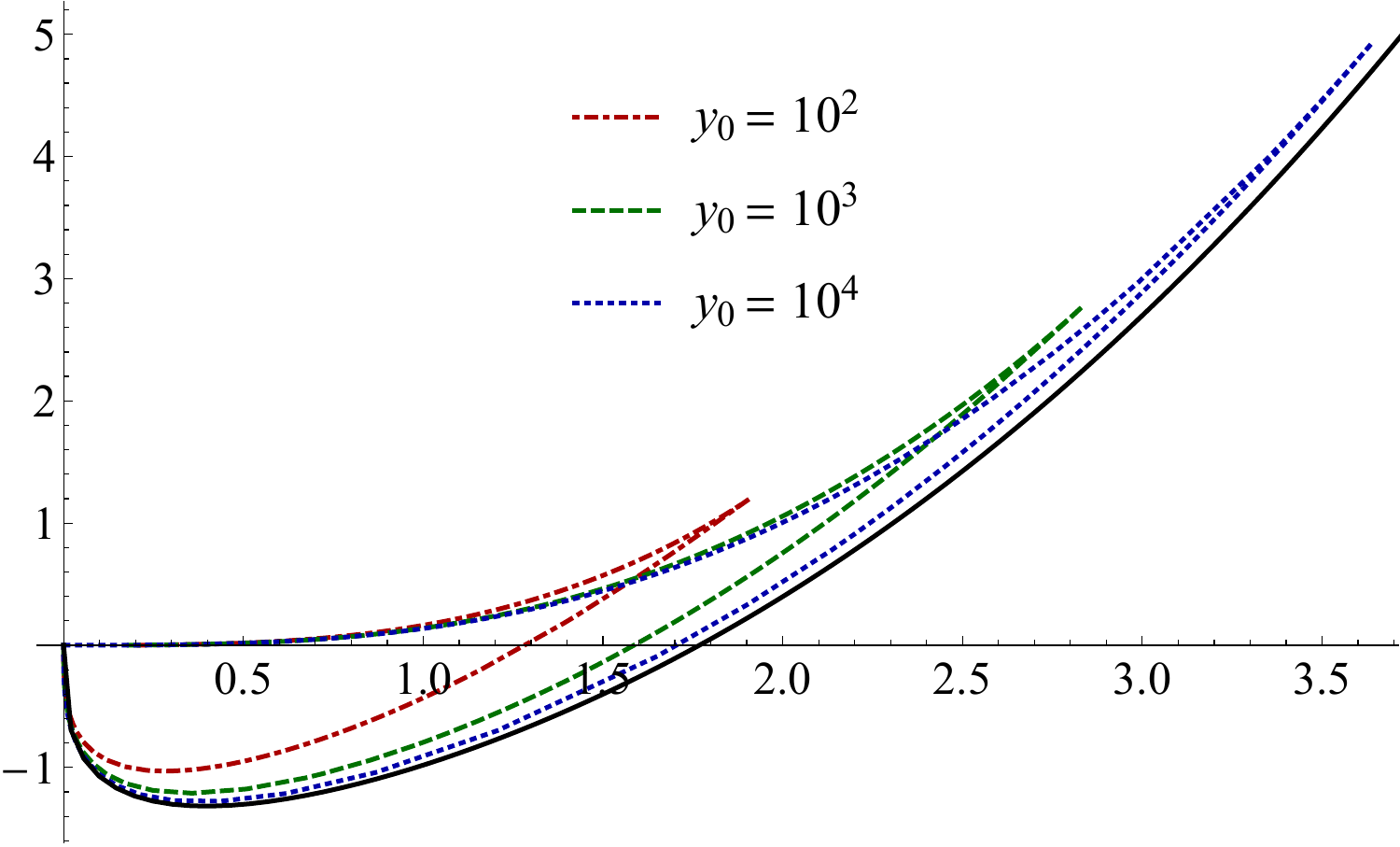} 
\put(-220,140){$V L / M_0L_0$}
\put(-30,6){$L/L_0$}
\end{subfigure}
\caption{\small (Left) Quark-antiquark potential for several $\Bplus$ solutions and for the $\Binf$ solution, shown as a continuous black curve. $M_0$ and $L_0$ are given in \eqq{MM} and \eqq{norm}, respectively. The red, dashed horizontal line at the bottom of the plot corresponds to the value of $VL$ for the OP fixed point. (Right) Quark-antiquark potential for several $\Bminus$ solutions with $Q_k$ scaled as $Q_k^2=\rho_0^2/2y_0$, as dictated by \eqq{rho0}, where $\rho_0$ is the scale of the $\Bconf$ solution, whose quark-antiquark potential is shown as a continuous black curve.}\label{WLfig}
\end{center}
\end{figure}

\begin{figure}[t]
\begin{center}
\begin{subfigure}{.45\textwidth}
\includegraphics[width=\textwidth]{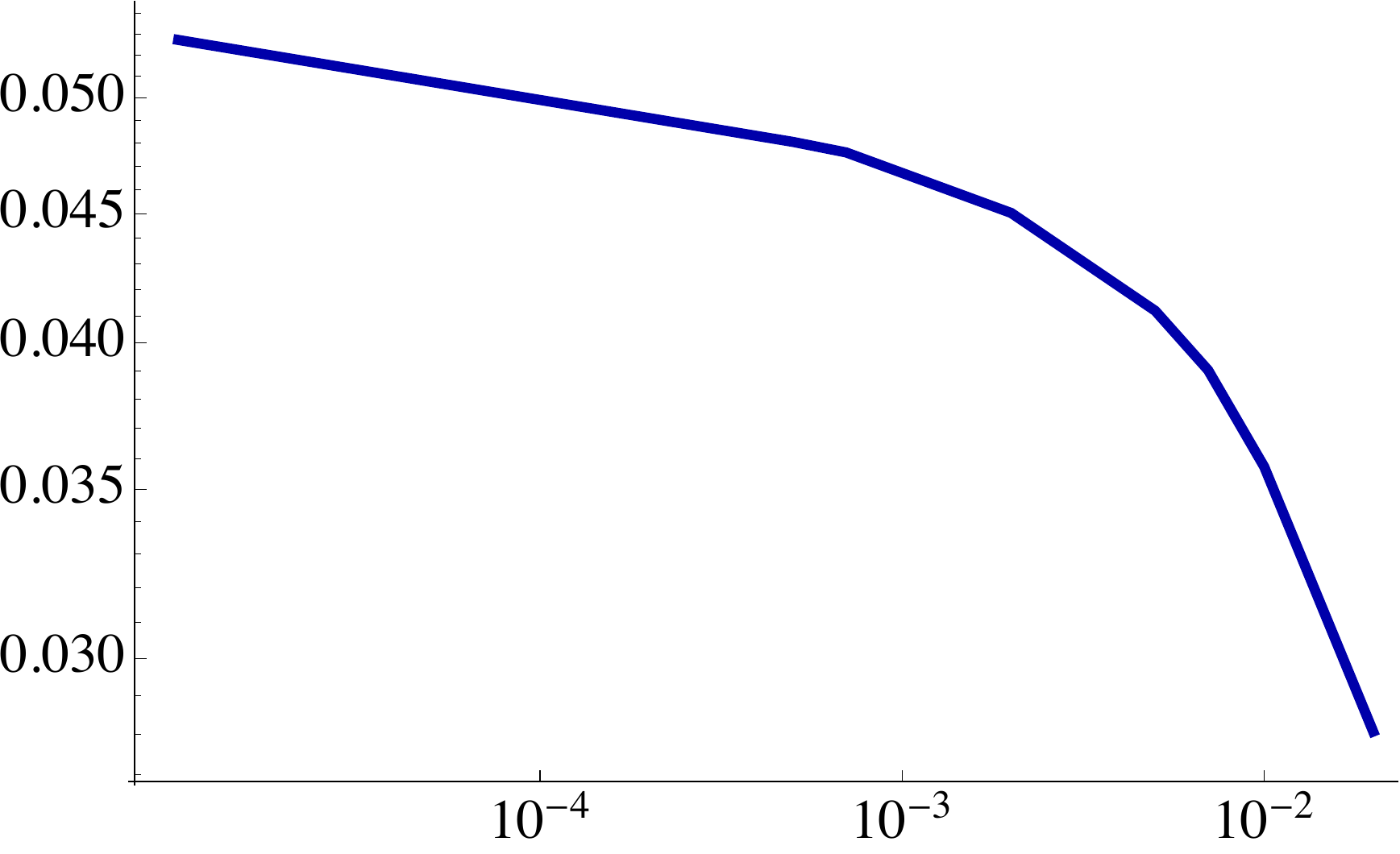} 
\put(-220,140){$M_* L_*' / M_0L_0$}
\put(-110,-15){$y_0+1$}
\end{subfigure}
\hfill
\begin{subfigure}{.45\textwidth}
\includegraphics[width=\textwidth]{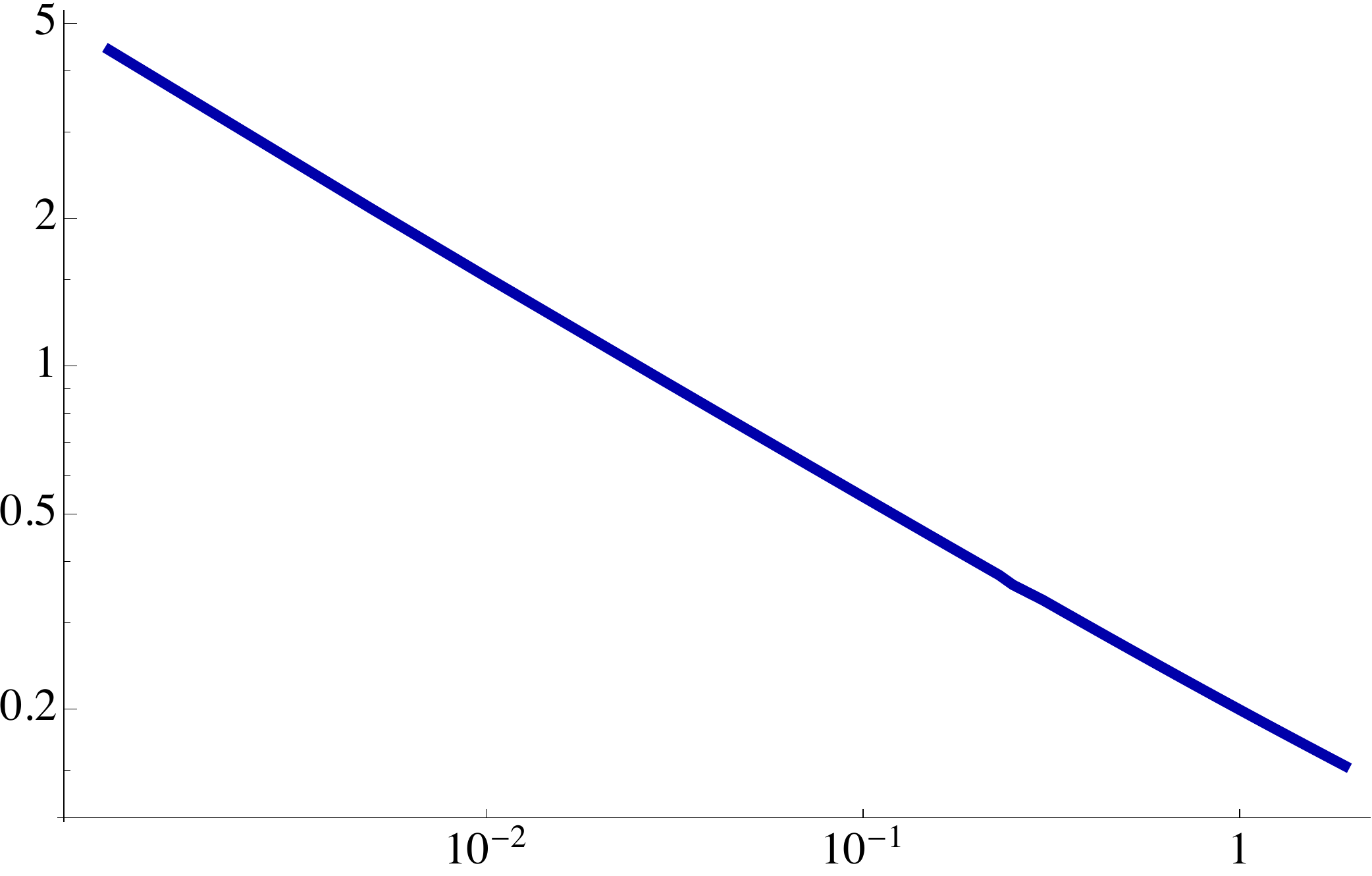} 
\put(-140,150){$M_* L_* / M_0L_0$}
\put(-110,-15){$y_0+1$}
\end{subfigure}
\caption{\small Comparison of the emergent IR scales extracted from the behavior of the dilaton and of the quark-antiquark potential.}
\label{illus}
\end{center}
\end{figure}

The results for the quark-antiquark potential $V$ as a function of the separation $L$ for the $\Bplus$ and $\Bminus$ solutions are shown in Figs.~\ref{WLfig}(left) and \ref{WLfig}(right), respectively, where $M_0$ is given by \eqq{MM} and  
\be
L_0 = \frac{\left(4q_c^2+3Q_c|Q_k|\right)^{1/2}}{|Q_k|^2} = 
\frac{6\pi N \sqrt{\bar M^2+2kN}}{\lambda \, k^2}  \,.
\label{norm}
\ee
The behavior of these curves can be understood as follows. In the UV, i.e.~in the limit $L\to 0$, the behavior is the same for all curves, since it is dictated by their common D2-brane asymptotics, which implies $V L\sim - L^{1/3}$. Thus, as $L$ begins to increase from zero, the curves first go down ($VL$ becomes more negative) until they reach a turning point and start going up. After this point all curves except for the one corresponding to $\Binf$ reach $V=0$ and cross the horizontal axis at a separation that we call $L_*$. When this happens  the disconnected configuration becomes energetically preferred. In the case of $\Binf$ the product $VL$ asymptotically approaches a negative constant  corresponding to the OP fixed point, as expected form the fact that this is the endpoint of the  $\Binf$ flow. In this case the preferred configuration is always the connected one. We see from Fig.~\ref{WLfig}(left) that curves with $y_0$ closer and closer to $-1$ become flatter and flatter and cross the horizontal axis at a larger and larger $L_*$. These curves cross the OP horizontal line at a smaller value $L_*' < L_*$. In this way we see that two IR mass scales emerge for flows that come close to the OP fixed point: 
\be
M_L = 1/L_* \sac M_L' = 1/L_*' \,.
\ee
In \Fig{illus} we compare these scales to the scale $M_*$ that we determined in \Fig{dil} from the behavior of the dilaton. Although it is difficult to push the numerics to arbitrarily small values of $y_0+1$, the plots in the figure suggest that, in the limit $y_0 \to -1$, $M_L'$ is of the same order as $M_*$ (at least over a large range of values of $y_0$), whereas $M_L / M_*$ goes to zero. In other words, the theory first deviates from the quasi-conformal behavior at the scale $M_* \sim M_L'$ but the membrane  becomes disconnected at a lower scale $M_L < M_*$.  

In the opposite limit, as $y_0\to\infty$, we see from \Fig{WLfig}(right) that the curves approach that of the $\Bconf$ solution, as expected from the discussion in \Sec{quasiconfining}.  This suggests that these theories exhibit quasi-confining dynamics. We will come back to this point in \Sec{disc}. 

At distances $L$ slightly larger than $L_*$ all curves but $\Bconf$ reach a cusp and ``turn back'', thus making the plots multivalued. The reason for the cusp is that, as  the penetration depth of the hanging membrane  inside the bulk increases beyond the point corresponding to the cusp,  $L$ begins to decrease.  This kind of behavior also appears in e.g.~calculations of the quark-antiquark potential in solutions with horizons, i.e.~in gauge theories at non-zero temperature. As in those cases, the part of the curve beyond the cusp is always energetically disfavored.

\section{Discussion}
\label{disc}
The solutions that we have studied in this paper provide a counterexample to the expectation that holographic duals of gauge theories with a mass gap also exhibit confinement. The key point is that, geometrically, this intuition is based on thinking of the quark-antiquark potential in string theory, where it is computed by a hanging string. In this case the smooth capping off of the geometry associated to the mass gap, together with charge conservation, which prevents the string from becoming disconnected, lead to a linear quark-antiquark potential at large distances, as illustrated in \Fig{string}. In our examples, however, the ten-dimensional description is singular and one must resort to  eleven-dimensional  M-theory in order to have a regular description. In this context the potential is computed from a hanging membrane which is allowed to become disconnected because the  M-theory circle on which  it  is wrapped  shrinks to zero size in the IR, as shown in \Fig{membrane}. This cuts off the linear growth of the potential at long distances. A necessary  condition for the consistency of these arguments is that the ten- and eleven-dimensional solutions are not simultaneously reliable in the IR, since otherwise they would have led to contradictory conclusions regarding the presence of confinement. Note also that the quark-antiquark potential we computed was a simplification because we did not allow the circle wrapped by the membrane to vary inside the ${\rm S}^7$. Nevertheless we expect that it captures the correct qualitative physics. 

Both the arguments of \Sec{quasi} and the quark-antiquark potential of \Sec{potential} suggest that the solutions we have discussed exhibit  quasi-conformal and quasi-confining behavior in the appropriate limits. The latter deserves some further discussion. Indeed, looking at \Fig{WLfig}(right) we see that the curves corresponding to solutions with $y_0 \to \infty$ approach the curve for the $\Bconf$ metric.  The latter shows the expected linear behavior $V\sim L$ at asymptotically large distances, which appears as a parabola in the figure since we are plotting  $VL \sim L^2$. Although the curves with large $y_0$ reproduce this behavior, we must remember that the connected configuration ceases to be preferred once the curves cross the horizontal  axis at $L=L_*$. Therefore, strictly speaking, the potential becomes constant at distances beyond this crossing point. However, the transition from the connected to the disconnected configuration is a semiclassical one, since it requires a fluctuation of the membrane in such a way that the circle transitions from non-zero to zero size. We expect that these fluctuations are exponentially suppressed provided the size of the circle is large in Planck units. For $\Bminus$ solutions  this can be achieved up to  arbitrarily long distance scales by taking $y_0$  and the appropriate charges to be large enough. For these solutions  connected configurations with $VL >0$ can be arbitrarily long lived, thus leading to an effective confining potential up to distance scales longer than  $L_*$. 

Interestingly, some of the features discussed above are  shared by four-dimensional Quantum Chromodynamics (QCD) if we think of the quark masses as adjustable parameters. Indeed, if all quarks are massive then QCD exhibits a mass gap but no confinement, since the growth of the potential between an external quark-antiquark pair is cut off by the breaking of the flux tube caused by the nucleation of a dynamical quark-antiquark pair. However, if the mass of the quarks is much larger than the QCD scale, then these nucleation is exponentially suppressed and the potential is effectively confining  over a large range of distances. 

It would be interesting to explore the analogies between the solutions that we have described and QCD further. For this purpose it would be useful to construct a four-dimensional analogue of our solutions. We leave these issues for future work.


\section*{Acknowledgements}

We are grateful to Javier Tarr\'\i o for collaboration at an early stage of this work. We thank him and Akikazu Hashimoto for comments on the manuscript.  We also acknowledge helpful discussions with  Yago Bea, Alex Buchel, Davide Cassani and  Christiana Pantelidou.  We are supported by grants MEC FPA2013-46570-C2-1-P, MEC FPA2013-46570-C2-2-P, MDM-2014-0369 of ICCUB, 2014-SGR-104, 2014-SGR-1474, CPAN CSD2007-00042 Consolider-Ingenio 2010, and ERC Starting Grant HoloLHC-306605.

\appendix

\section{Reduced Action}

In this section we write the ansatz for reducing type IIA on the coset ${\rm Sp}(2)/{\rm U}(2)$. We will keep only the scalars. For the full $\mathcal{N}=2$ reduced supergravity see \cite{Cassani}. We take the string-frame metric to be
\begin{equation}
 \dd s_s^2\,=\,e^{\Phi/2}\left(e^{-2U-4V}\dd s_4^2+e^{2U}\,\frac14\left[\left(E^1\right)^2+\left(E^2\right)^2\right]+e^{2V}\,\frac12\dd\Omega_4^2\right)
\end{equation}
in such a way that the NK point corresponds to $U=V=0$. The dilaton $\Phi$ only depends on the external coordinates and the fluxes take the same form as in (\ref{all}) with $a_X=0$ for consistency. 

We find that all the type IIA equations of motion and Bianchi identities  are verified if the equations of motion deduced from the following action are satisfied
\begin{eqnarray}
S&=&\frac{1}{2\kappa_4^2}\,\int\,\left[R*1-G_{ij}\dd\phi^i\wedge*\dd\phi^j-\mathcal{V}*1\right]\nonumber\\[2mm]
&=&\frac{1}{2\kappa_4^2}\,\int\,\left[R*1-\frac12\left(\dd\Phi\right)^2-4\left(\dd U\right)^2-12\left(\dd V\right)^2-8\dd U\cdot\dd V-4e^{-4V-\Phi}\left(\dd b_J\right)^2\right.\nonumber\\[2mm]
&&\qquad\qquad-8e^{-4U-\Phi}\left(\dd b_X\right)^2-32e^{-2U-4V+\Phi/2}\left(\dd a_J\right)^2-\mathcal{V}*1\bigg]\,,
\end{eqnarray}
with the potential
\begin{eqnarray}
\mathcal{V}&=&128\,e^{- 6 U -12 V-\Phi/2} \left[Q_c + 4 a_J \left(b_J + b_X\right) +Q_kb_J\left(b_J-2b_X\right)+2 q_c \left(b_X - b_J\right)\right]^2 \nonumber\\[2mm]
&+&  32 \left(b_J + b_X\right)^2 e^{-4 U - 8 V - \Phi} + 64 \left[2 a_J +Q_k\left(b_J-b_X\right) -  q_c\right]^2 e^{-6 U - 8 V + \Phi/2} \nonumber\\[2mm]
&+& 32 \left(2 a_J -Q_kb_J + q_c\right)^2 e^{-2 U - 12 V + \Phi/2} +4Q_k^2 e^{-2 U - 8 V + 3\Phi/2 } \nonumber\\[2mm]
&+&8Q_k^2 e^{-6 U - 4 V + 3\Phi/2}- 24 e^{-2 U - 6 V} -  8 e^{-4 U - 4 V} + 2 e^{-8 V} \,. 
\end{eqnarray}
Remarkably, this potential follows from the superpotential
\begin{eqnarray}
\mathcal{W}_{\pm}&=&e^{-4 V} + 2 e^{-2U -2 V} + Q_k\,e^{-3 U - 2 V + 3\Phi/4} - Q_k\,e^{-U - 4 V + 3\Phi/4} \\[2mm]
&\pm& 4 e^{-3 U - 6 V - \Phi/4} \left[Q_c + 4 a_J \left( b_J + b_X\right) +Q_k b_J \left(b_J-2b_X\right)+2q_c \left(b_X - b_J\right) \right]\nonumber
\end{eqnarray}
through the usual relation
\begin{equation}
\mathcal{V}\,=\,4\,G^{ij}\partial_i\mathcal{W}\partial_j\mathcal{W}-6\,\mathcal{W}^2\,.
\end{equation}
Using the domain-wall  ansatz
\begin{equation}
\dd s_4^2\,=\,e^{2A}\dd x_{1,2}^2+\dd \rho^2
\end{equation}
the resulting set of BPS equations coincide with (\ref{BPSsystem}), (\ref{Fluxes}) and (\ref{BPSwarp}) upon the identifications
\begin{eqnarray}
e^{\Phi}&=&h^{1/4}\,e^{\Lambda}\,,\nonumber\\[2mm]
e^{2U}&=&4\,h^{3/8}\, e^{2g-\Lambda/2}\,,\nonumber\\[2mm]
e^{2V}&=&2\,h^{3/8}\, e^{2f-\Lambda/2}\,,\nonumber\\[2mm]
e^{2A}&=&16\,h^{1/2}\, e^{4f+2g-2\Lambda}\,,
\end{eqnarray}
together with the change of radial coordinate
\begin{equation}
\dd\rho\,=\,4\,h^{3/4}\,e^{2f+g-\Lambda}\dd r\,.
\end{equation}
It is a consistent truncation to fix $b_J=-b_X=(2q_c)/(3Q_k)$ and $a_J=-q_c/6$. The potential thus admits two supersymmetric extrema at
\begin{equation}
e^{2U}\,=\,2\,e^{2V}\,=\,\frac{4\left(4q_c^2-3Q_cQ_k\right)^{3/8}}{3^{3/4} Q_k^{1/4}}\,,\qquad\qquad e^{\Phi}\,=\,\frac{\left(4q_c^2-3Q_cQ_k\right)^{1/4}}{3^{1/2}Q_k^{3/2}}\,,
\end{equation}
for positive $Q_k$, and 
\begin{equation}
e^{2U}\,=\,\frac25\,e^{2V}\,=\,\frac{4\left(4q_c^2+3Q_c|Q_k|\right)^{3/8}}{3^{9/8}\times 5^{3/8}\times|Q_k|^{1/4}}\,,\qquad\qquad e^{\Phi}\,=\,\frac{\left(4q_c^2+3Q_c|Q_k|\right)^{1/4}}{3^{3/4}\times5^{1/4}\times|Q_k|^{3/2}}\,,
\end{equation}
for negative $Q_k$. The uplifts of these extrema correspond to the ABJM and OP fixed points, and  thus preserve $\mathcal{N}=6$ and $\mathcal{N}=1$ supersymmetry, respectively. They are also extrema of the superpotential with the $+$ and $-$ signs in the last term respectively. 
The spectrum of scalar fluctuations together with the dimension of their dual operators at the OP fixed point is shown  in Table~\ref{scalarspectrum}.

\begin{table}[h]
\resizebox{0.95\textwidth}{!}{\begin{minipage}{\textwidth}
\begin{center}
\begin{tabular}{lcc}
\hline
\rowcolor{gray}	Mass Eigenstate&$ \qquad m^2L^2$&$\qquad\Delta$\\[6pt]
\hline
$12\delta U+24\delta V+\delta\Phi$&$\qquad18$&$\qquad6$\\[6pt]
\rowcolor{gray}	$4\delta a_J-Q_k\left(2\delta b_J+5\delta b_X\right)$&$\qquad10$&$\qquad5$\\[6pt]
$3\delta\Phi-4\delta U$&$\qquad\frac{52}{9}$&$\qquad\frac{13}{3}$\\[6pt]
\rowcolor{gray}	$\delta a_J+Q_k\delta b_J$&$\qquad\frac{10}{9}$&$\qquad\frac{10}{3}$\\[6pt]
$2\delta a_J-Q_k\left(\delta b_J-15\delta b_X\right)$&$\qquad-\frac{8}{9}$&$\qquad\frac{8}{3}$\\[6pt]
\rowcolor{gray}	$8\delta U-12\delta V+3\delta\Phi$&$\qquad-\frac{20}{9}$&$\qquad\frac{5}{3}\left(\frac43\right)$\\[6pt]
\hline
\end{tabular}
\end{center}
\caption{\small \small Spectrum of scalars around the $\mathcal{N}=1$ supersymmetric AdS$_4$ solution.}
\label{scalarspectrum}
\end{minipage} }
\end{table}

\section{Spin(7) Manifolds from Tri-Sasakian Geometry}

In the main text we have considered a family of flows built around the seven-sphere as the internal manifold, and therefore whose dual gauge theory is related to ABJM or its supersymmetric reduced version, the OP theory. In this appendix we will show that what was secretly exploited in \cite{Cvetic2, Cvetic3} to construct Spin(7)-manifolds is the tri-Sasakian structure on ${\rm S}^7$. In this way we will extend the results of this paper  to other of $\mathcal{N}=3$ gauge theories conjectured to be dual to these seven-dimensional tri-Sasakian manifolds. 

There exist several equivalent definitions of tri-Sasakian manifolds. For our purposes the most convenient one is that they admit a triplet of one-forms $\eta^I$ and a triplet of two-forms $J^I$ so that
\begin{equation}\label{extdev}
\dd\eta^I\,=\,2J^I-\epsilon^I{}_{LM}\eta^L\wedge\eta^M\,,\qquad\qquad\qquad\dd J^I\,=\,2\epsilon^I{}_{LM}J^L\wedge\eta^M\,,
\end{equation}
with $\iota_{\xi_I}J^K=0$, being $\xi_I$ the vector dual to $\eta^I$. Indeed, the triplet of Killing vectors $\xi_I$, generating the algebra of SO(3), defines an SU(2) or SO(3) foliation over a quaternionic K\"ahler base (QK), whose triplet of almost complex structures is precisely $J^I$. In this way, the tri-Sasakian metric is given by  
\begin{equation}
\dd s_{\textrm{\tiny 3S}}^2\,=\,\dd s^2\left({\rm QK}_4\right)+\left(\eta^1\right)^2+\left(\eta^2\right)^2+\left(\eta^3\right)^2\,.
\end{equation}
It can also be shown that 
\begin{eqnarray}\label{Hduals}
J^L\wedge J^M&=&2\delta^{LM}\,\omega_{\textrm{\tiny QK}}\,\nonumber\\[2mm]
*\left(J^I\wedge\eta^{L_1}\wedge\dots\wedge\eta^{L_n}\right)&=&\frac{1}{(3-n)!}\epsilon^{L_1\dots L_n}{}_{L_{n+1}\dots L_{3}}\,J^I\wedge\eta^{L_{n+1}}\wedge\dots\wedge\eta^{L_3}
\end{eqnarray}
with $\omega_{\textrm{\tiny QK}}$ the volume form of the quaternionic K\"ahler base. 

We propose now the following eight-dimensional metrics constructed from these squashed tri-Sasakian manifolds
\begin{equation}\label{3Sanstaz}
\dd s_8^2\,=\,\dd t^2+4a^2\left[\left(\eta^1\right)^2+\left(\eta^2\right)^2\right]+4b^2\,\left(\eta^3\right)^2+4c^2\,\dd s^2\left({\rm QK}_4\right)\,,
\end{equation}
where $a$, $b$ and $c$ are functions of $t$ chosen in order to facilitate  the comparison with \cite{Cvetic2}. Furthermore, on this space we define the four-form
\begin{eqnarray}
\Psi&=&16 c^4\,\omega_{\textrm{\tiny QK}}+8a^2c^2\,\epsilon_{LMN}\eta^L\wedge\eta^M\wedge J^N+8b\,\dd t\wedge\left(b^2\,\eta^1\wedge\eta^2\wedge\eta^3+c^2\,\eta^I\wedge J^I\right)\,.
\end{eqnarray}
Using the properties listed in Eqs.~(\ref{extdev}) and (\ref{Hduals}) it can be seen that this four-form is closed and self-dual provided the following system of equations is satisfied
\begin{eqnarray}\label{3Ssystem}
a'&=&1-\frac{b}{2a}-\frac{a^2}{c^2}\,,\nonumber\\[2mm]
b'&=&\frac{b^2}{2a^2}-\frac{b^2}{c^2}\,,\\[2mm]
c'&=&\frac{a}{c}+\frac{b}{2c}\,.\nonumber
\end{eqnarray}
This set of equations coincides with the one studied in \cite{Cvetic2} and ensures Ricci-flatness of the metric. Moreover, it is possible to show that $\Psi$ has the symmetries of the octonionic structure constants and thus is suitable for being the Cayley form defining Spin(7) manifolds. When the tri-Sasakian foliation is chosen to be ${\rm S}^7$ one recovers the results of \cite{Cvetic2} and, through the identifications (\ref{abc}), everything discussed in the bulk of the paper.

By picking the solution
\begin{equation}
a\,=\,b\,=\,\frac{3}{10}\,t \sac c\,=\,\frac{3}{2\sqrt{5}}\,t\,,
\end{equation}
we get the eight-dimensional metric
\begin{equation}
\dd s_8^2\,=\,\dd t^2+\frac95\,t^2\left\{\dd s^2\left({\rm QK}_{4}\right)+\frac{1}{5}\left[\left(\eta^1\right)^2+\left(\eta^2\right)^2+\left(\eta^3\right)^2\right]\right\}\,,
\end{equation}
which is the Spin(7)-cone over the weak-${\rm G}_2$ metric that every (squashed) tri-Sasakian manifold admits and whose dual would be an $\mathcal{N}=1$ CFT. In the case of the seven-sphere this is the OP point. 

For other solutions it remains to be seen if the ansatz (\ref{3Sanstaz}), together with a particular solution to (\ref{3Ssystem}), defines globally a regular metric. In any case, the solutions analogous to $\mathbb{A}_8$ in \cite{Cvetic2} would describe a flow from a $\mathcal{N}=1$ SYM-like quiver gauge theory to the $\mathcal{N}=3$ IR fixed points with tri-Sasakian duals. 

Let us now give some details about the geometries appearing in type IIA after reduction. On every tri-Sasakian there is an ${\rm S}^2$-worth of Sasaki--Einstein (SE) structures, each one defined by picking a ${\rm U}(1)\subset{\rm SO}(3)$ isometry through the Killing vector $\xi=\alpha^I\xi_I$, with $\alpha^I\alpha^J\delta_{IJ}=1$, and characterized by the forms 
\begin{equation}
\eta\,=\,\alpha_I\eta^I\,,\qquad\qquad\qquad J\,=\,\alpha_M\left(J^M-\frac{1}{2}\epsilon^M{}_{NL}\eta^N\wedge\eta^L\right)\,,
\end{equation}
and the metric
\begin{equation}
\dd s^2_{\textrm{\tiny SE}}\,=\,\dd s^2\left({\rm KE}_6\right)+\left(\eta\right)^2\,,
\end{equation}
that is, a U(1) fibration over a six-dimensional K\"ahler--Einstein (KE) base with metric $\dd s^2\left({\rm KE}_6\right)$. The K\"ahler form on the base is precisely $J$. In the case of the seven-sphere, this base is $\mathbb{CP}^3$.  Since on the SE manifold $\dd\eta\,=\,2J$ we can write $\dd\psi+C_1$ with $\dd C_1\,=\,2J$. It follows that,  
after reduction along $\psi$, the type IIA two-form flux will be proportional to the K\"ahler form and the internal metric will be $\dd s^2\left({\rm KE}_6\right)$. It turns out that the reduced internal space admits also a nearly K\"ahler metric with almost complex structure 
\begin{equation}
J_{\textrm{\tiny NK}}\,=\,\alpha_M\left(J^M+\frac{1}{4}\epsilon^M{}_{NL}\eta^N\wedge\eta^L\right)\,,
\end{equation}
which corresponds, in this paper, to the squashed $\mathbb{CP}^3$ appearing as the internal geometry in the UV of the D2-branes.


\end{document}